# Haptic Empathy: Investigating Individual Differences in Affective Haptic Communications

Yulan Ju
Keio University Graduate School of Media Design
Yokohama, Japan
yulan-ju@keio.jp

Xiaru Meng
Keio University Graduate School of Media Design
Yokohama, Japan
xiarumeng2019@outlook.com

Harunobu Taguchi
Keio University Graduate School of Media Design
Yokohama, Japan
h.tag@kmd.keio.ac.jp

Tamil Selvan Gunasekaran
Empathic Computing Lab, Auckland Bioengineering Institute
University of Auckland
Auckland, New Zealand
tg469@aucklanduni.ac.nz

Matthias Hoppe
Keio University Graduate School of Media Design
Yokohama, Japan
JSPS International Research Fellow
Tokyo, Japan
matthias.hoppe@kmd.keio.ac.jp

Hironori Ishikawa
NTT DOCOMO
Tokyo, Japan
ishikawahiron@nttdocomo.com

Yoshihiro Tanaka
Nagoya Institute of Technology
Nagoya, Japan
Inamori Research Institute for Science
Kyoto, Japan
tanaka.yoshihiro@nitech.ac.jp

Yun Suen Pai
School of Computer Science
University of Auckland
Auckland, New Zealand
Empathic Computing Laboratory
Auckland Bioengineering Institute,
University of Auckland
Auckland, New Zealand
yun.suen.pai@auckland.ac.nz

Kouta Minamizawa
Keio University Graduate School of Media Design
Yokohama, Japan
kouta@kmd.keio.ac.jp

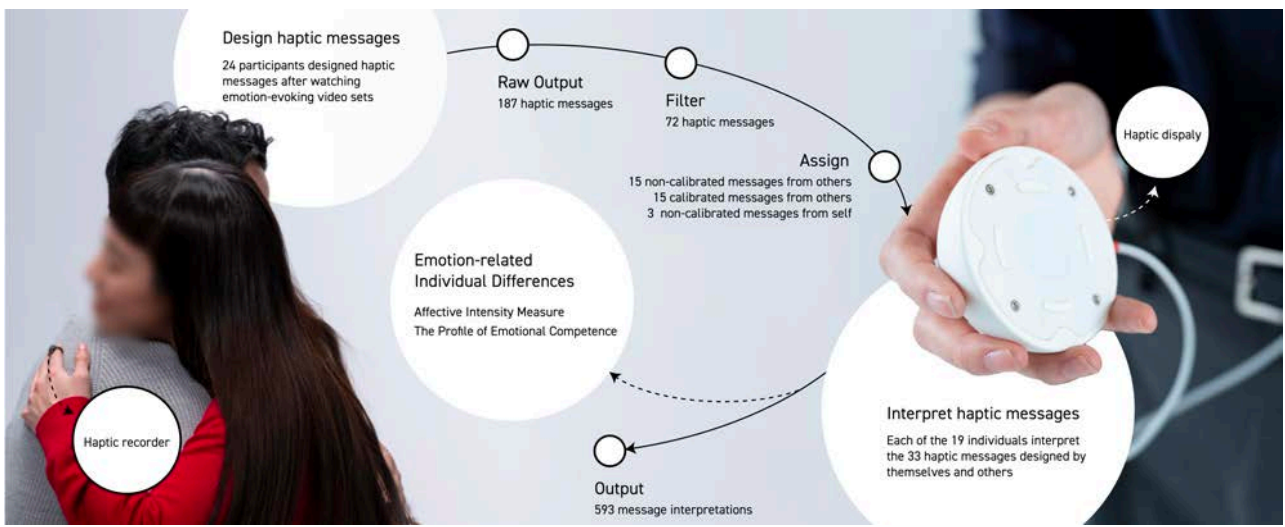

**Figure 1: The workflow detailing the process of the study detailed in this work, from designing the haptic messages to the interpretation, output, and subsequent analysis of said affective haptic messages. A total of 187 haptic messages were generated by 24 participants based on emotion-evoking video sets, which were then filtered down to 72 haptic messages. Afterwards, 19 participants were each assigned 33 calibrated and non-calibrated haptic messages designed by themselves and others to interpret, yielding a total of 593 interpretations, which were then analyzed for emotion-related individual differences.**

## Abstract

Nowadays, touch remains essential for emotional conveyance and interpersonal communication as more interactions are mediated remotely. While many studies have discussed the effectiveness of using haptics to communicate emotions, incorporating affect into haptic design still faces challenges due to individual user tactile acuity and preferences. We assessed the conveying of emotions using a two-channel haptic display, emphasizing individual differences. First, 24 participants generated 187 haptic messages reflecting their immediate sentiments after watching 8 emotionally charged film clips. Afterwards, 19 participants were asked to identify emotions from haptic messages designed by themselves and others, yielding 593 samples. Our findings suggest potential links between haptic message decoding ability and emotional traits, particularly Emotional Competence (EC) and Affect Intensity Measure (AIM). Additionally, qualitative analysis revealed three strategies participants used to create touch messages: perceptive, empathetic, and metaphorical expression.

## 1 Introduction

Social interaction is a multifaceted exchange of information; multiple modalities and cues ranging from subtle to overt forms of communication. Verbal exchanges allow for a direct and orderly exchange of information about a topic through words [17], while non-verbal exchanges focus more on conveying mannerisms, thoughts, behaviors, and other more subtle cues, usually through body language [5]. Previous studies have shown that conveying emotional contexts is one of the main purposes of non-verbal communications between individuals [65]. Additionally, among the many non-verbal methods to convey emotions such as facial expressions, gestures, tone of voice, prosody, and so on, touch has proven to be able to intensify interpersonal communication and convey intimate emotions [20, 40, 71].

When examining haptic communication systems, one finds that they generally fall into two primary classifications: those oriented towards task completion and those aimed at conveying emotions. The majority of academic exploration has been directed towards understanding task-focused, non-affective touch communication, such as navigation, reminders, or systems like Tactile Braille. However, integrating emotion into haptic design presents its own set of challenges. Recent research [79, 99] has made progress in documenting affective responses to synthetic tactile stimuli, but obstacles still remain. Namely, the fact that synthetic signals are used as vibrotactile stimuli means the relationship between affect and haptics is often documented in a discrete manner. This discrete approach can limit the understanding of the nuanced interplay between emotions and tactile sensations. Adding to the complexity, the existing literature also suggests that individual differences exist in the perception of haptics [64, 76, 78]. Most research tends to evaluate user preferences after the design process, exemplified by the quantitative assessment of haptic perception and interactions on a scale. Only a few studies [66] have discussed the reason behind users' predilection for specific haptic sensations or interactions. Decisions taken throughout research formulation, design framework, and evaluation may inadvertently direct the emphasis towards the technological apparatus rather than the intended user demographic [93]. Furthermore, only limited information is available regarding users' mental framework for vibrations, the qualitative and affective characteristics they associate with these signals, and the fundamental semantic structure underlying their meaning.

Taking all this into account, we are particularly interested in understanding **user-centric variability** in expressing emotions and understanding emotions through **non-synthetic vibrotactile stimuli**. The central focus of this research is to explore the variability in haptic message design and interpretation, emphasizing the role of individual differences in emotional traits. This approach will contribute to a more nuanced understanding of Haptic Empathy and its potential applications in enhancing emotional communication through haptics. To explore the impact of interpersonal variations on the design and interpretation of affective haptic messages, this study seeks to answer the following research questions:

**RQ1**: How do individuals intuitively design touch-based vibrations to express different emotions?

**RQ2**: How do emotional traits such as Emotional Competence (EC) and Affect Intensity Measure (AIM) influence the design and interpretation of haptic messages?

**RQ3**: What design strategies can optimize haptic interfaces for conveying emotional meaning, taking into account interpersonal differences?

We conducted our exploration based on an existing work by Ju et al. [40], where they define Haptic Empathy as the ability to extract and encode emotional meaning in the tactile aspects of human-to-human interactions or convey and understand emotional meanings through haptics. Similarly, our investigation to establish the individual differences between users can be broken down into two procedures: designing affective haptic messages and interpreting them. During the design process, participants would watch affective videos, followed by encoding a haptic message using natural touch. For the interpretation phase, the participants were required to replay the previously encoded messages and provide individual ratings regarding the emotions that they thought the messages were trying to convey. From our analysis, we found that users, when designing haptic messages in distinct emotional states, can effectively communicate certain emotional information. However, this communication standard is highly individualistic. Notably, there exists a correlation between proficiency in designing haptic signals and the skills to interpret them. Furthermore, the capability to decode these haptic messages appears to be potentially associated with specific emotional traits, particularly Emotional Competence (EC) and Affect Intensity Measure (AIM). Our study offers key contributions to the field of affective haptic communication, emphasizing interpersonal variability and emotional traits:

- **Refinement of Haptic Empathy:** We advance the concept of Haptic Empathy by emphasizing interpersonal differences in emotional expression and interpretation through haptics, capturing a wide array of intuitive methods people use when employing touch-based vibration to convey emotions.
- **Empirical Evidence on Interpersonal Variation:** Our empirical findings demonstrate how emotional competencies and affect intensity influence the coding and decoding of haptic messages. This highlights the importance of accounting for individual emotional traits in haptic communication systems.
- **Guidelines for Personalized Haptic Design:** Based on our findings, we propose design principles that accommodate individual differences in emotional perception and communication, enhancing the effectiveness of haptic systems in conveying emotional meaning. This work lays the foundation for future research into personalized haptic empathy systems that adapt to users' emotional traits, facilitating intuitive and effective emotional communication through touch.

## 2 Related Work
## 2.1 Touch in Social Communication

Our sense of touch plays a crucial role in physical interaction with our environment and those around us. It serves as a primary channel for non-verbal communication, conveys deep emotions, and is essential for our physical and emotional health. In today's digital era, much of human social interaction is mediated through technology. To fully grasp the potential of haptic interfaces for social touch, it is essential to first understand the various impacts of non-mediated human-to-human touch.

*2.1.1 Non-mediated Human to Human Social Touch.* Touch, as a vital component in human interaction and communication, takes various forms in our daily lives, including greetings, intimate exchanges, and corrective actions [92]. The existing body of research points to the significant role social touch plays in various aspects of the human experience, specifically with regards to physical and emotional well-being, attachment and bonding, attitudinal and behavioral change, and communication of affect [14, 38]. While this breadth of influence that touch has underscores the importance of designing haptic interfaces, our research places a heavier focus on communicating emotions through touch.

Research in multimodal communication indicates that touch not only stands alone as a form of communication but also enhances other verbal and non-verbal methods by amplifying the intensity of emotional displays from our face and voice [46]. On the other hand, Hertenstein et al. [32, 33] explored that touch communicates not just the intensity but also the distinctness of each emotion. They found that people are capable of discriminating between at least nine distinct emotions - including anger, disgust, fear, gratitude, happiness, love, sadness and sympathy - through being touched on the arm or body by a stranger, even in the absence of visual cues. Similarly, a recent work [97] also investigated how slight distinctions in skin-to-skin contact influence both the recognition of cued emotional messages (e.g., anger, sympathy) and the rating of emotional content (i.e., arousal, valence). These studies have identified strong associations between emotions and specific touch gestures, kinematics of the gestures (pressure intensity, speed, acceleration and duration), and skin-to-skin contact attributes (velocity, depth, and area).

*2.1.2 Mediated Social Touch.* To meet the demands of long-distance interpersonal communication in modern times, research in the field of affective haptics is focused on the development of interfaces that can transmit social touch from one user to another or from a machine to a user. This "ability of one actor to touch another actor over a distance by means of tactile or kinesthetic feedback technology" has been defined as mediated social touch [31]. In terms of facilitating affective haptic communication between two people, a wide range of systems have been developed to represent various touch interactions such as kissing [70], hugging [15, 59, 89, 91], squeezing [68], handholding [29, 90], and more, using everything from hand strokes to thermal signals. Beyond direct methods, indirect approaches like avatars in virtual environments have also been used. For instance, haptic-jacket systems [37] can be used in virtual worlds to allow users to exchange touch sensations like pats or hugs. Further, social touch can be used to make virtual agents perceived as more human-like and blur the boundary between avatars and agents [36].

A large body of research is also working on improving and enriching the communications between a human user and a virtual entity. To be more specific, researchers are trying to develop systems that can feel, understand, and respond to social touch signals. Often, systems that are capable of producing social touch are only geared towards improving the emotional experience, such as an affective haptic jacket embedded in a movie scene or game [3, 4, 49] or a device that maps the emotions evoked by music into vibrations for a hearing-impaired audience [43, 56]. Apart from the scripted and one-way social touch cues used in the above-mentioned examples, applications in human-computer interaction are now increasingly incorporating intelligent agents to enhance the social dimension of interactions [92]. Several studies seem to indicate that interacting with social robots can offer benefits similar to those of social companions and therapists [10, 69, 94]. In addition to communicating affective cues using facial expressions, gestures, and speech, an increasing number of these social agents [30, 81, 101], whether physically embodied as robots or represented as virtual characters on-screen, are employing touch technology for human interaction. The touch signal is not only transmitted through a medium but is also both generated and interpreted by an electronic system, rather than directly by a human being.

## 2.2 Design and Effectiveness of Affective Haptic Interfaces

Often, the goal of mediated social touch is to mimic the feel of touch as it naturally occurs in daily life. Achieving this authenticity requires a blend of natural interaction methods and high-fidelity tactile displays. Current research in mediated social touch incorporates gesture recognition technology to enable natural gesture-based interactions and combines various tactile technologies to recreate a broad range of tactile sensations as accurately as possible. In the following sections, we will delve into the advancements of current technologies from these two angles and specifically discuss the vibration-based technologies that are central to our research.

*2.2.1 Recognizing Affective Touch.* Building on the touch gestures and kinematics identified earlier, researchers in the field of affective haptics have primarily utilized these insights to develop software capable of automatically recognizing users' emotional states from the way they touch a display, a keyboard, or a robot. For example, projects like the "Corpus of Social Touch" (CoST) [41] and "Human-Animal Affective Robot Touch" (HAART) [24] have contributed data sets that document various touch behaviors, including patting, scratching, stroking, and squeezing. Using these two data sets as an evaluation, Y. Gaus et al. [27] have proposed an ensemble of machine learning methods to discriminate among these types of touch behaviors. This technology is widely used in interactions with affective social robots, such as pet robots [2, 30, 101]. It is worth mentioning that besides identifying affective gestures and serving as an interactive trigger, studies that combined gesture recognition information with direct affect recognition found that good gesture recognition improves the effectiveness of touch-based affect recognition systems [2].

*2.2.2 Haptic Modalities.* Current haptic research focuses on developing pervasive, unobtrusive, and natural interfaces for high-fidelity haptic rendering, utilizing various actuation technologies and materials for skin stimulation based on physical parameters. Commonly employed technologies include vibrotactile, thermal, force, and contactless methods, often integrated to expand the bandwidth of haptic information [19, 20, 72]. Some studies incorporate multiple technologies in a single interface for more comprehensive haptic experiences [1, 4, 58, 66, 95]. Despite advancements, haptic technology has yet to achieve the sophistication necessary to fully replicate the complexity of real human touch.

Users, however, have shown the capacity to infer nuanced affective information from simplified haptic signals like vibration, even with current technological limitations [31, 40, 74]. Vibrotactile feedback, a popular haptic method, simulates texture by generating vibrations and is favored for its affordability, simple design, and ease of control. Vibrotactile actuators are typically integrated into handheld devices or wearables, allowing for a wide range of applications [11, 18, 34, 39, 67]. Research has explored the engineering properties of vibration and their effect on affective perception. For instance, Yliopisto et al. encoded emotional cues by modifying frequency and duration parameters [98], while Yoo et al. demonstrated a relationship between four physical parameters—amplitude, frequency, duration, and envelope—and the valence and arousal of vibrotactile patterns [99]. Seifi et al. further aided haptic designers with VibViz, a database offering design variables and taxonomies to optimize vibrotactile cues [79].

To enhance expressivity, tactor arrays have been developed, offering rich spatiotemporal vibrotactile patterns across larger skin areas. Initial efforts focused on basic directional patterns [86], while more advanced systems mimic real-world sensations like a snake slithering up an arm [80]. Such systems have been used to convey affective information through vibrotactile displays, such as the Haptic Face Display (HFD), which uses vibration motors to display emotional responses [57]. Similarly, a study by Cang et al. explored the potential of vibrotactile animations using tactor arrays along the forearm to transmit affect-embedded tactile messages [12].

Despite these advances, authentic replication of human touch remains a significant challenge. The complex social nature of touch and technological limitations restrict the range of touch experiences that can be accurately mimicked. Most devices simulate touch using vibrations, pressure, or temperature but fail to capture the full nuances of human touch. This gap is especially evident in remote communication technologies, where the recognition and reproduction of touch—particularly the role of the touch initiator—are underexplored [66, 68]. While physical parameters are often the focus in vibration-based interfaces, there is a need for more expressive and realistic methods to initiate and convey touch, especially in mediated social interactions. Developing technologies that support natural, varied touch initiation could significantly improve remote haptic communication.

## 2.3 Human Dynamics in Affective Haptic Communication

While the above-mentioned works demonstrate that haptics can communicate distinct emotions effectively, they also reveal that the overall performance of the systems is largely dependent on a variety of factors, such as gender, relationship status, cultural background, familiarity, social status, and so on. In response to these challenges, several studies have emphasized the need to personalize haptic signals to enhance their utility and user adoption. For instance, in evaluating the efficiency of a haptic stress regulation device, Bequet et al. [7] observed that participants with higher scores in neurotic and extroverted personality traits experienced more subjective relaxation using the device, which highlights the importance of customizing haptic regulation methods to individual characteristics such as personality traits. Another example is the work of Seifi et al. [77], where they manipulated the perception of vibrations along three emotion dimensions: agitation, liveliness, and strangeness. Their goal was to provide users with the capability to "directly manipulate" haptic signals. While their findings show promise for developing new vibrotactile interfaces, it is important to note that their work primarily involved adjusting already manipulable physical parameters, in which these synthetic vibrotactile signals may not fully capture the nuanced aspects of tactile perception.

Furthermore, it is clear that the ability to infer emotional context from haptic stimuli depends on far more than the physical properties of said haptic stimuli. Therefore, in addition to characterizing the link between the engineering parameters of vibrations and emotional attributes, certain research has concentrated on human factors and the particularities of human perception. Taking a look at the literature, we can see that three categories of individual variances have been documented [75]:

- Sensing and perception: Variations can exist in the sensitivity and ability to distinguish signals of mechanoreceptors among individuals, resulting in disparities in tactile acuity, detection thresholds, and difference detection [52, 62]. These disparities are particularly noticeable for subtle sensations like programmable friction and may have an influence on the overall range of perceptible tactile experiences. Research by Schaefer et al. [73] has found a link between tactile performance and the personality trait of empathy. Specifically, they noted that the empathic concern subscale is positively associated with performance in tasks requiring tactile acuity.
- Tactile processing and memory: Individuals display a varying degree of proficiency in processing and learning tactile stimuli [23, 35, 44, 82]. Early studies using OPTACON (OPtical to TActile CONverter), an electromagnetic device that allows those with visual impairments to read text that hasn't been transcribed into Braille, established two groups of "learners" and "non-learners" for tactile matching tasks [35], suggesting a clear difference in learning trajectories [83]. Similar results were observed in studies that utilized a variable friction interface [50]. Other studies suggest variance in the extent to which individuals rely on touch for information gathering and noted that haptic processing ability is a skill that can be improved over time, which was backed up by the finding that those with visual impairments usually develop outstanding tactile processing abilities [25].
- Meaning mapping and preference: Individuals often feel the need to connect abstract signals to their intended meanings. In the absence of universally recognized cultural associations regarding haptics, the process of assigning meanings to abstract haptic signals relies heavily on personal experiences and individual thresholds for sense-making. Differences in these thresholds suggest the presence of distinct, personalized frameworks for mapping meaning to haptics [26, 50].

While previous research already demonstrated the power of touch as an effective communication channel, studies mainly focus on mapping the physical parameters of haptics to emotions. As such most of these studies explore the possibility of conveying and interpreting stereotypical affective messages and touch gestures from the perspective of technical capabilities. However, there is still a limited understanding of the human cognitive style and emotional intelligence influencing the creation and interpretation of natural tactile interactions.

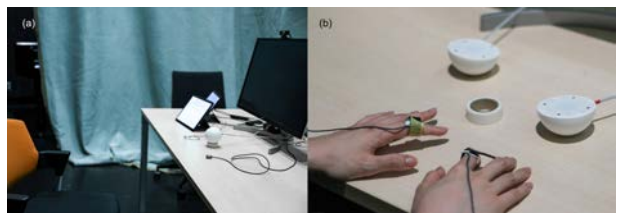

Figure 2: Experimental setup: (a) Haptic display with a tablet to assess participants' sensitivity levels; (b) two PVDF films are outfitted on the index fingers of both hands for tactile recording.

This paper aims to bridge this knowledge gap by providing a deeper understanding of these processes in specific contexts and semi-ecological settings. Our approach combines qualitative studies with controlled recognition studies. Instead of directing participants to display predetermined emotional meanings, we immerse them in an emotional context and then ask them to express these emotions through touch. By allowing users to engage with the system freely and creatively without any constraints, our research uncovers insights into how people intuitively use haptic technology without prior training. These findings are versatile, applicable to diverse contexts, and can inform the design of similar interfaces in the future.

## 3 Methodology

All experiments were conducted in accordance with the organization's ethical policy regarding human participant testing. All participants gave their written informed consent prior to their inclusion in the study. The experiment was arranged in a sound-proofed, temperature-regulated environment. All materials were placed on a desk (see Fig.2 (a)) containing a haptic display used to share haptic messages, two tactile recording devices (PVDF films [87, 88]), a display used to present the guideline and stimuli, and a pair of noise-cancelling headphones. Two cameras were used for high-definition facial and full-body recordings. A large curtain separated the participants from the researchers, creating a semi-enclosed space.

### 3.1 Haptic Display

*3.1.1 Building the Prototype.* The wearable haptic device that was developed in collaboration with NTT Docomo, consists of two half-spheres, each with a diameter of 63 mm, housing the haptic actuators (Foster Electric Company, #639897). These hemispheres are designed to magnetically combine into a single smooth sphere, offering both aesthetic appeal and functional versatility. The device interfaces with computers via USB, functioning as a sound output device. The design is driven by key user-centric requirements: it is lightweight and portable, ensuring that it can be held for 30 minutes without causing user fatigue. It features emotionally neutral colors to avoid any visual cues that might influence the outcomes of the experiment. Furthermore, the device delivers high-fidelity vibrations for a realistic haptic experience and is intuitively designed for ease of use, requiring no prior explanation. This device aligns with the partner company's branding and design ethos, integrating consumer needs and technical expertise from our internal team. The choice to focus on hand-based interaction is grounded in the sensitivity of the hands and social acceptability for touch, enhancing the user's comfort and data interpretation capabilities.

*3.1.2 Haptic Recording.* To ensure unrestricted freedom of touch in our study, we used PVDF films as haptic recording sensors considering that skin vibrations are a key factor in the perception of roughness [63]. These sensors took on the form of a belt, consisting of a PVDF film attached to a velcro strap with a rigid base for length adjustment. The sensor was positioned on the finger pad, specifically between the distal interphalangeal and proximal interphalangeal joints, as illustrated in Fig.2 (b). The sensor is designed to detect skin-propagated vibrations, which are generated from the mechanical interaction between the fingertip and an object, but it does not respond to the overall movement of the finger, which allows users to touch objects with their bare fingers. To capture the haptic data, we recorded signals from both the left and right hands using a Portacapture X8 field recorder.

For creating prerecorded haptic content, we used Premiere Pro video editing software to adjust recorded data. This adjustment involved processes like equalization and pitch shifting, facilitating users to experience vibrotactile sensations that are in sync with video and audio. The content, as depicted in Appendix A, showcases the potential of haptic technology in various applications, including skill transmission in performance and traditional crafts, as well as in telemedicine.

*3.1.3 Calibration system.* The system calibrates the intensity of tactile stimuli based on the sensation level among individuals. The sensation level is calculated by dividing the amplitude of the tactile stimuli by the detection threshold. In this system, detection thresholds for three frequencies of 50, 200, and 400 Hz are measured as representative characteristics as humans have different sensitivities to frequency [28], and the detection threshold from 0 to 2000 Hz is estimated. The transfer function is obtained from the sender's and receiver's detection thresholds for each frequency to make the sensation level of tactile stimuli between the sender and receiver, and the amplitude of the tactile stimuli recorded is modulated through the short Fourier Transform and inverse Fourier Transform by using the transfer function obtained [45].

## 3.2 Message Meaning Phase: Stimuli Selection

Film clips are an important tool for evoking emotional responses in a laboratory setting. While other various emotionally charged visual stimuli, namely pictures, exist, film clips have proven to be superior in eliciting emotions for longer periods of time at both the subjective and physiological levels. Notably, the majority of these film clips are cross-modal, engaging both the auditory and visual senses [13]. However, considering the interplay between vibrotactile stimuli and other modalities such as audio, there's an emphasis on ensuring that one modality does not overshadow the others. To create a set of realistic, informal meanings, enabling participants to design affective haptic messages without undue external influence, we used 8 carefully selected 40-second film clips without audio from the EMDB data sets [13], designed to induce specific emotions from the participants (see Table 2). This database was validated the effectiveness of the 40-second clips using psychophysiological sensing (Skin Conductance Level (SCL) and heart rate). Drawing from the valence and arousal findings in their publication, we classified the film clips into four categories: high valence high arousal (HVHA), high valence low arousal (HVLA), low valence high arousal (LVHA), and low valence low arousal (LVLA). Each category comprises two distinct film clips.

## 3.3 Self-assessment forms

Primary methods to capture affective responses include self-report measures and biometric recordings. Research suggests that self-reports demonstrate a heightened sensitivity in detecting subtle affective variations between stimuli compared to biometric recordings. Therefore, self-report measures are often favored when analyzing smaller, relative differences between stimuli [84]. In our study, we embedded the self-assessment manikin (SAM) [8] in the questionnaires for both the design and interpretation phases (see Appendix B), to capture valence and arousal levels. In terms of valence, a continuous scale comprised of nine manikins illustrated feelings ranging from very negative (scored 1) to very positive (scored 9). For arousal, a second set of nine manikins depicted feelings of very calm (scored 1) to very excited (scored 9). Images of these manikins corresponded to the nine response options. To maintain consistency in self-assessment, participants were asked to answer digital forms in both phases using an iPad that we provided.

## 3.4 Design Phase

An overview of the entire experimental procedure can be seen in the flowchart in Fig.3. A total of 24 participants (self-reported as 12 male, 12 female; aged $\mu$ = 28.0 years, $\sigma$ = 4.0) from varying ethnicities (Caucasian, African-American, Asian, and Latino or Hispanic) were recruited . Participants generated haptic messages based on the short film clips in Table 2 within an isolated setting. At the beginning of the procedure, the participants were informed that some of the film clips might be graphic in nature as some of the clips contained fictional depictions of violence and/or sexual content and were reminded that they were free to exit the experiment at any time. Each session took around 60 minutes, and each participant was compensated with a small honorarium of $15.

*3.4.1 Understanding the Haptic Display.* First, each participant was given a tutorial on the haptic experience sharing system [85]. The procedure began with instructions to hold the haptic presentation device as shown in Fig.4(a), while we collected their tactile sensitivity to each frequency band by delivering sine wave vibrations of gradually increasing intensity in three frequency bands of 50Hz, 200Hz, and 400Hz. Based on the tactile sensitivity, a sensitivity curve was estimated and the delivered tactile information was frequency equalized accordingly. Participants then watched a 3-minute-long video with pre-designed haptic feedback embedded in the video (the contents can be found in Appendix A) to gain an initial understanding of haptic technology and the capabilities of current haptic displays. We asked participants to hold the two hemispheres separately (see Fig.4(b)) and suggested imitating the hand movements in the video (see Fig.4(c)) for a better, more immersive experience.

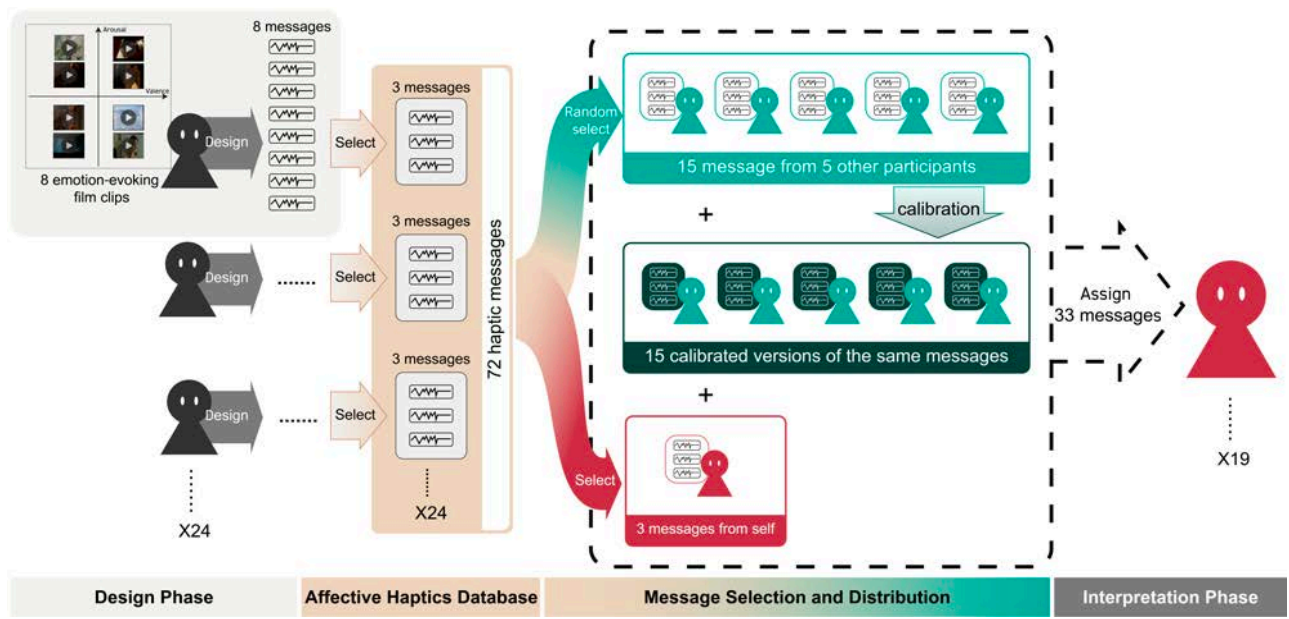

Figure 3: A flowchart of the experimental procedures detailed in this work. During the design phase of the experiment, a total of 24 participants were shown 8 emotion-evoking film clips and asked to generate a haptic message for each clip. A total of 72 messages (3 per participant) were selected for further study, based on self-reported confidence levels and balanced across different film clip types. For the interpretation phase, each returning participant was assigned 33 haptic messages to interpret, consisting of 15 randomly-selected non-calibrated haptic messages from 5 other participants, 15 calibrated versions of those same messages, and 3 non-calibrated messages made my themselves.

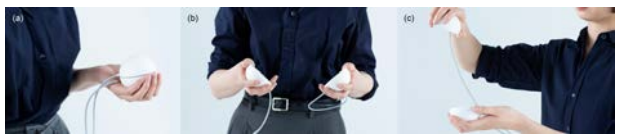

Figure 4: Gesture Guide - Three ways of operating the haptic display as instructed to the participants during the tutorial: (a) Holding it in one hand to collect tactile sensitivity information; (b) Holding the two hemispheres separately in each hand while they watched a 3-minute demo video with pre-designed haptic feedback so they can gain an initial understanding of the haptic technology and its capability; and (c) Imitating the movements in the said video for better immersion.

After the demonstration, we put PVDF sensors on both index fingers of the participants' hands for them to familiarize themselves with haptic message recording and the device's capability. Participants were encouraged to explore various materials using different tactile actions, such as tapping, rubbing, and pressing. The practice trials lasted for 3 minutes.

*3.4.2 Participant Message Creation.* To design the haptic message, each participant was brought into an isolated space to work independently and was asked to put on noise-cancelling headphones to prevent any external distractions. We informed the participants of the detailed process before the formal encoding session and guided each step through the formal recording via text on a monitor facing the participant. Within one session, each participant designed haptic messages for 8 film clips from Table 2 in random order, ending with a self-assessment form. Specifically, we allocated time for: (a) a 40-second movie clip to elicit a specific emotional state in the participant; (b) recording of a 5 to 10-second haptic message (with a 10-second countdown on the display as a reminder to the participant) and then allowed the participants to check the recording and record again if they deemed it necessary; (c) 3 minutes to complete the self-assessment (45 seconds) of arousal, valence, dominance and selection of basic emotions (see Fig.10 (design phase)), with the rest of the time being used for breaks to reduce the impact of the emotion fluctuations generated by the previous emotion expression on the latter sample. After 8 trials, the recording session ended. At the end of this session, we asked participants how they felt about the task and whether they used any strategies to express their emotions via the haptic device.

*3.4.3 Affective Haptics Database.* From the design phase, a total of 187 unique haptic messages were collected, which are considered the raw output. From this set, further selection was carried out as shown in Fig. 3 to achieve a well-curated dataset for analysis and interpretation. Specifically, three haptic messages were selected from each participant based on their confidence levels and balanced across different film clip types to ensure a diverse range of stimuli for interpretation and analysis. The selection process prioritized high-confidence-level messages from each participant, focusing on those with the highest ratings on a scale from 1 (lowest) to 5 (highest). Additionally, the distribution of messages was carefully balanced to ensure an equal number of messages across the four quadrants representing different emotional and arousal levels: LVHA, LVLA, HVHA, and HVLA. This approach resulted in a total of 72 haptic messages with an average confidence score of 3.63, which forms the Affective Haptics Database, with detailed information presented in the table 1.

**Table 1: Distribution of haptic messages across film clips, showing the count, maximum, minimum, and mean confidence levels for each clip.**

| Clip | Count | Conf. Max | Conf. Min | Conf. Mean |
|---|---|---|---|---|
| Clip 1 (LVHA) | 8 | 5 | 1 | 3.00 |
| Clip 2 (LVHA) | 10 | 5 | 2 | 3.80 |
| Clip 3 (HVHA) | 9 | 5 | 2 | 3.78 |
| Clip 4 (HVHA) | 9 | 5 | 1 | 3.22 |
| Clip 5 (HVLA) | 10 | 5 | 2 | 3.27 |
| Clip 6 (LVLA) | 10 | 4 | 4 | 4.00 |
| Clip 7 (LVLA) | 7 | 5 | 2 | 3.57 |
| Clip 8 (HVLA) | 9 | 5 | 3 | 4.33 |

## 3.5 Interpretation Phase

To assess how accurately these haptic messages could be interpreted, we invited all the participants who had participated in the first message-generating session back for another phase of the study for interpretation one week later. By involving the same individuals in both the design and interpretation stages, we aimed to explore the relationship between each participant's design strategies and their interpretation abilities [12, 40]. Of the original 24 participants, 19 returned for the interpretation phase. In total N = 19 individual participants (self-reported as 11 male, 8 female; aged $\mu$ = 29.0 years, $\sigma$ = 4) were played a set of haptic messages. They were informed that the set of messages was picked randomly from the design phase. Each session took about 60 minutes in total. Each participant was compensated with a small honorarium of $10 USD after the experiment.

*3.5.1 Message Selection and Distribution.* Each of the returning participants were given a set of 33 haptic messages, consisting of 15 randomly-selected non-calibrated messages from 5 other participants, 15 calibrated versions of those same messages, and 3 non-calibrated messages made by themselves (see Fig.3 (Message Selection and Distribution)).

*3.5.2 Procedure.* Before the interpretation session began, participants were given instructions for the rating scale used in the procedure (see Fig. 10 (b)). Participants wore noise-cancelling headphones throughout the entire session, and all instructions were given in text form via a display in front of the participant. In order to avoid fatigue caused by high concentration over a long period of time, which could affect decoding ability, we structured the session into three trials, each consisting of 11 haptic messages, with a 5-minute break in between each trial. Participants were asked to fill out the rating scale after each haptic message, and they were allowed to replay the message multiple times if they were unsure of their answer. Finally, a semi-structured interview was conducted after all interpretation was completed.

## 3.6 Emotion-related Individual Differences

To investigate whether the capacity to interpret affective haptic cues correlates with certain emotion-related abilities, we profiled our participants using Affect Intensity Measure (AIM) [47] and the Profile of Emotional Competence (PEC) [9]. Affect intensity is a stable individual difference characteristic defined in terms of the typical strength of an individual's responsiveness [48]. This intensity is consistent across various emotional categories, suggesting an inherent temperament related to emotional reactivity and variation. Emotional Competence (EC) describes an individual's proficiency in the identification, understanding, expression, regulation and use their own and others' emotions. This ability significantly impacts how individuals adapt to and interact with their surroundings. We used the profile of emotional competence (PEC) scale in our study to assess our participants from intrapersonal (pertaining to one's own emotions) and interpersonal (related to the emotions of others) dimensions for each competency.

## 4 Analysis and Results

## 4.1 Pre-processing of the Dataset

Before proceeding with the complete analysis, we excluded a subset of interpreted data compromised by errors during data logging operations. This adjustment left us with 593 data points. Given the relatively small sample size of messages designed by the participants, analyzing them separately could limit the statistical power of the analysis. Additionally, since these messages were designed by the participants, their interpretation inherently relies on the individual's own skin sensitivity, making calibration status unlikely to have a meaningful influence on how these self-designed messages are interpreted. Duplicating and re-labeling these messages as calibrated allowed for a more balanced representation between the calibrated and non-calibrated pools, ensuring a reliable basis for further analysis. This adjustment resulted in a total of 650 data points.

## 4.2 Analysis of the Effectiveness of the Stimuli

We initially evaluated if the film clips were able to effectively elicit the intended feelings in the participants, essentially determining whether they designed haptic messages while in the intended specific emotional states. As illustrated in Table 2, the valence and arousal scores reported by participants closely aligned with the mean scores from the database. The inter-class correlation coefficient (ICC) for valence was remarkably high at 0.982 (p < .001). Likewise, the arousal dimension had a significant ICC of 0.912 (p = .002). This implied that participants' feelings were correctly elicited. The results implied that the selected film clips effectively served as stimuli for this study.

Table 2: Details of the videos selected for the experiment. The mean values and standard deviation of valence and arousal that were reported in the experiment are shown in comparison to the values reported in the database; we are able to see that the selected clips were able to immerse the participants into the intended emotions.

| No. | No.in EMDB | Film clip content | Clip description | Video Type | Valence mean (SD) | | Arousal mean (SD) | |
|---|---|---|---|---|---|---|---|---|
| | | | | | Database | Experiment | Database | Experiment |
| 1 | 1006 | Horror | Cannibal Holocaust | LVHA | 1.98 (1.50) | 2 (1.41) | 7.37 (1.88) | 6.43 (2.19) |
| 2 | 1009 | Horror | Midnight Meat Train | LVHA | 1.83 (1.24) | 2.22 (1.41) | 6.88 (1.70) | 5.83 (2.21) |
| 3 | 2002 | Erotic Couples | 9 songs | HVHA | 7.15 (1.50) | 6.38 (2.02) | 5.99 (1.84) | 5.38 (1.93) |
| 4 | 2009 | Erotic Couples | Diary of a Nymphomaniac | HVHA | 6.54 (1.74) | 6.13 (2.11) | 6.11 (1.85) | 5.63 (2.12) |
| 5 | 4007 | Social positive interactions | Diary of a Nymphomaniac | HVLA | 6.58 (1.44) | 7.38 (1.28) | 3.38 (1.79) | 4.79 (2.34) |
| 6 | 3007 | Social negative interactions | American Beauty | LVLA | 3.20 (1.19) | 3.65 (1.07) | 3.58 (1.84) | 4.09 (1.59) |
| 7 | 3001 | Social negative interactions | The descent | LVLA | 3.04 (1.67) | 3.78 (1.41) | 4.79 (1.42) | 3.26 (1.36) |
| 8 | 5002 | Scenery | Disney's Earth | HVLA | 5.68 (1.70) | 6.22 (1.76) | 2.51 (1.86) | 1.83 (0.94) |

### 4.3 Overview of the Dataset

To explore the similarities and differences in the emotional dimensions between the designer and interpreter of the haptic messages, we defined the absolute difference in valence scores between them as the "valence distance". If the valence distance equals 0, valence accuracy is termed "correct" with a chance level of 11.1%, and it is scored as 1. If the valence distance equals 1, the valence accuracy is considered "partially correct" with a chance level of 19.8%, and it is scored as 0.5. Finally, if the valence distance exceeds 1, valence accuracy is classified as "incorrect" with a chance level of 69.1% and is scored as 0 [60]. "Arousal distance" and "arousal accuracy" were defined and scored following the same method. (See Appendix E for detailed calculation methods used)

We initiated our analysis with a comprehensive evaluation of the gathered data. The haptic messages were first classified by their calibration status, calibrated and non-calibrated. Initially, the Spearman rank correlation coefficient, as well as the p-value, were calculated to examine the relationship between the designer and interpreter's ratings of valence and arousal (see Table 3(a)). For calibrated haptic messages, there was a weak positive correlation between the designer and interpreter's valence ($\rho$ = 0.235, p < .001) and a moderate positive relationship for arousal ($\rho$ = 0.397, p < .001). Both correlations are statistically significant. For the non-calibrated haptic message, a weak positive correlation was found between the designer's and interpreter's valence ($\rho$ = 0.199 (p < .001), and a moderate positive correlation was observed between the designer and interpreter's arousal ($\rho$ = 0.417, p < .001).

Additionally, the binomial test with exact Clopper-Pearson 95% CI (see Table 3(b)) were used to evaluate whether the observed proportions of "correct" and "partially correct" classifications for valence and arousal exceeded their respective chance levels (11.1% for "correct" and 19.8% for "partially correct"). For the calibration group, the accuracy for valence was low (p=0.102, 95% CI = 0.099, 0.179), failing to surpass the chance level. However, the proportion of partially correct responses was significantly higher than chance (p=0.009, 95% CI = 0.207, 0.307). Arousal responses in the calibration group demonstrated higher accuracy (p<0.001, 95% CI = 0.145, 0.235) and partially correct rates (p=0.003, 95% CI = 0.216, 0.317), both exceeding their respective chance levels. Similar trends were observed for the non-calibration group, with accuracy and partially correct rates for arousal consistently above chance level, while valence accuracy remained closer to chance levels. These results suggest that calibrated messages improve interpretative accuracy, particularly for arousal, while valence interpretations remain more challenging across both conditions.

Upon further classification of the haptic messages by their association with film clip types (HVHA, HVLA, LVHA, LVLA), we generated a heatmap (see Fig.5) to visualize the transition of SAM distribution from the intention behind the haptic message design to the interpretation of the messages. The heatmap revealed that both valence and arousal values tended to move closer to neutral after interpretation, suggesting that when transforming visual stimuli into haptic stimuli, the emotional intensity is diminished. Additionally, the graph illustrates the differences between calibrated and non-calibrated messages. After calibration, there is improved clustering for HVHA messages. SAM scores for LVLA and HVLA messages become more concentrated, while, in contrast, LVHA messages show greater dispersion in SAM scores after calibration. For a more detailed analysis, we derived specific metrics (see Table 4). For each set of haptic messages, we calculated the average valence or arousal score from the interpreters, denoted as "Value". Similarly, we established the mean values for "Distance" and "Accuracy". The Spearman rank correlation coefficient was employed to calculate the association between the "Value" scores selected by the designer and the interpreter for both valence and arousal dimensions. Specifically, HVLA calibrated messages demonstrated a moderate positive correlation in arousal between the designer and interpreter ($\rho$ = 0.497, p < .001), while HVLA non-calibrated messages also showed a moderate positive correlation ($\rho$ = 0.418, p < .001). Similarly, for LVLA non-calibrated messages, a moderate positive correlation in arousal was observed ($\rho$ = 0.541, p < .001). In the case of LVHA calibrated messages, a moderate positive relationship in arousal was also found($\rho$ = 0.399, p < .001). All these observed correlations were confirmed to be statistically significant.

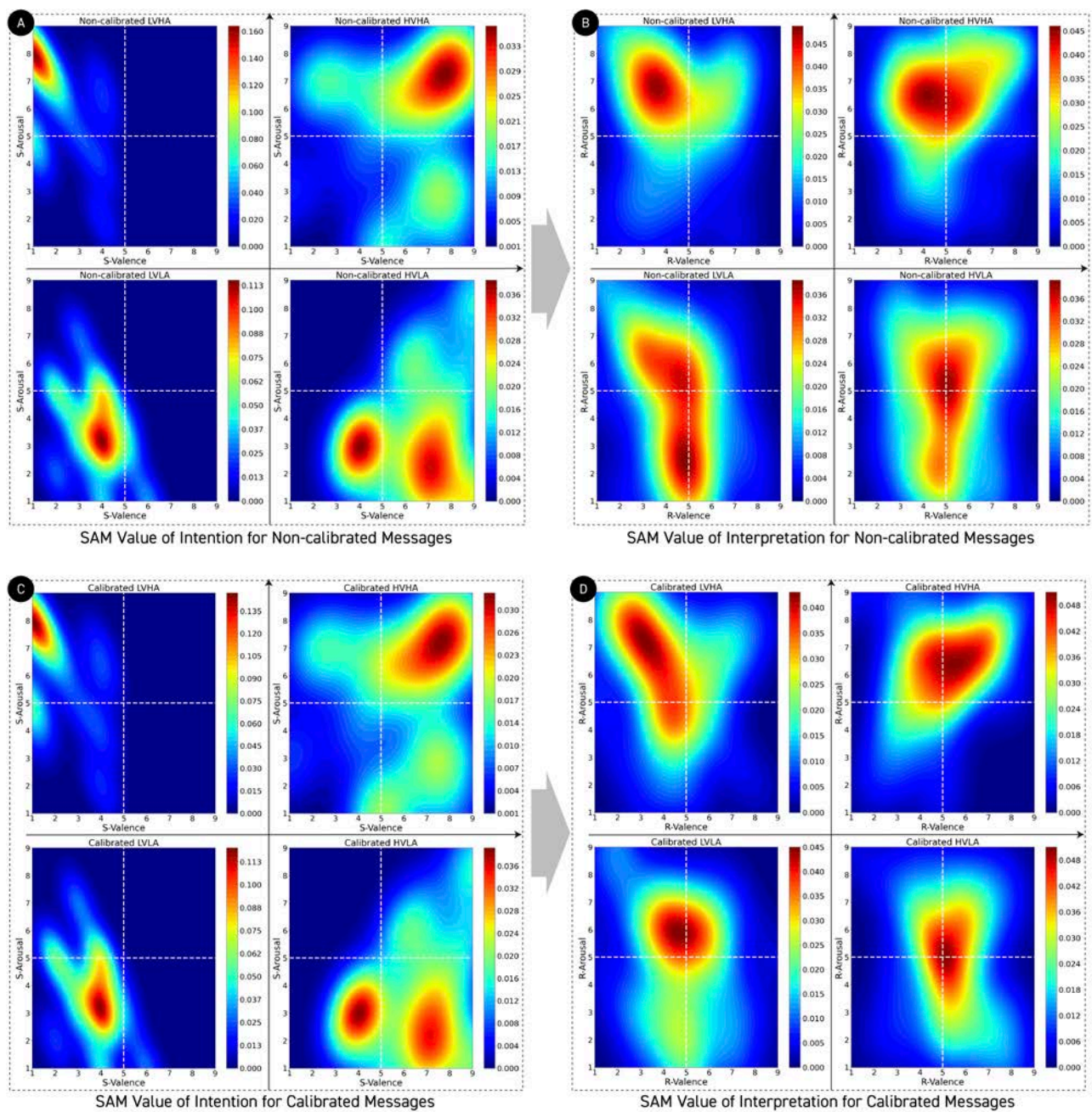

Figure 5: SAM distribution of haptic messages classified by association with film clip types (HVHA, HVLA, LVHA, LVLA) and calibration statements. The top two sets of heat maps (A and B) represent non-calibrated messages while the bottom two sets (C and D) represent calibrated ones. Heat maps on the left (A and C) represent the intention and the ones on the right (B and D) represent interpretation. Valence and arousal values showed a tendency to move closer to neutral post-interpretation, suggesting a diminished emotional intensity after transforming visual stimuli into haptic stimuli.

Table 3: Spearman Correlation and Clopper-Pearson Binomial Test Results

(a) Spearman Correlation Results. The table presents the Spearman correlation coefficients ($\rho$) and corresponding p-values for valence and arousal measures across Calibration and Non-Calibration groups.

| Group | Measure | Correlation ($\rho$) | p-value |
|---|---|---|---|
| Calibration | Valence | 0.235 | <.001 |
| | Arousal | 0.397 | <.001 |
| Non-Calibration | Valence | 0.199 | <.001 |
| | Arousal | 0.417 | <.001 |

(b) Clopper-Pearson Binomial Test Results for validation of the dataset collected. The table shows the proportion of correct and partially correct classifications with 95% confidence intervals (CI) for valence and arousal in both Calibration and Non-Calibration groups. It also includes a chance comparison to determine whether the proportions are statistically above chance level.

| Group | Measure | | Proportion (95% CI) | Chance Comparison |
|---|---|---|---|---|
| Calibration | Valence | Correct | 0.102 (0.099, 0.179) | Not above chance |
| | | Partially Correct | 0.207 (0.207, 0.307) | Above chance |
| | Arousal | Correct | 0.145 (0.145, 0.235) | Above chance |
| | | Partially Correct | 0.216 (0.216, 0.317) | Above chance |
| Non-Calibration | Valence | Correct | 0.160 (0.096, 0.170) | Not above chance |
| | | Partially Correct | 0.224 (0.224, 0.321) | Above chance |
| | Arousal | Correct | 0.137 (0.137, 0.221) | Above chance |
| | | Partially Correct | 0.246 (0.246, 0.346) | Above chance |

Table 4: Evaluation of haptic messages classified by association with film clip types (HVHA, HVLA, LVHA, LVLA) and calibration statements. Any accuracy rate presented in the table that exceeds 0.21 is regarded as being above the level of chance. All accuracy rates meet this criterion, except for the accuracy rate of interpreting the valence of both calibrated and non-calibrated LVHA type clips.

| Variables | | Value Mean(±SD) | Distance Mean(±SD) | Accuracy Mean(±SD) | Correlation rs (p) |
|---|---|---|---|---|---|
| **Valence** | | | | | |
| Calibration | HVHA | 4.95±1.68 | 2.16±1.66 | 0.30±0.39 | 0.103 (0.349) |
| | HVLA | 5.35±1.52 | 2.18±1.57 | 0.25±0.35 | -0.007 (0.950) |
| | LVHA | 4.18±1.72 | 2.61±1.71 | 0.19±0.34 | 0.038 (0.742) |
| | LVLA | 4.69±1.63 | 1.76±1.34 | 0.31±0.36 | -0.029 (0.813) |
| Non-Calibration | HVHA | 4.99±1.64 | 2.11±1.64 | 0.31±0.40 | 0.176 (0.094) |
| | HVLA | 5.00±1.55 | 2.36±1.69 | 0.24±0.36 | 0.042 (0.696) |
| | LVHA | 4.27±1.74 | 2.87±1.71 | 0.14±0.27 | -0.201 (0.064) |
| | LVLA | 4.37±1.57 | 1.47±1.04 | 0.38±0.35 | 0.241 (0.037) |
| **Arousal** | | | | | |
| Calibration | HVHA | 6.14±1.79 | 2.22±1.89 | 0.33±0.40 | 0.221 (0.042) |
| | HVLA | 4.89±2.11 | 2.13±1.66 | 0.30±0.39 | 0.497 (<.001) |
| | LVHA | 5.96±1.99 | 1.70±1.43 | 0.37±0.39 | 0.399 (<.001) |
| | LVLA | 4.70±2.20 | 1.99±1.41 | 0.27±0.37 | 0.356 (0.003) |
| Non-Calibration | HVHA | 6.22±1.78 | 2.14±1.73 | 0.32±0.37 | 0.237 (0.023) |
| | HVLA | 4.71±2.23 | 2.21±1.69 | 0.29±0.37 | 0.418 (<.001) |
| | LVHA | 5.93±2.03 | 1.74±1.48 | 0.36±0.39 | 0.355 (0.001) |
| | LVLA | 4.57±2.37 | 1.79±1.47 | 0.33±0.41 | 0.541 (<.001) |

## 4.4 Individual Interpretation Accuracy

In the subsequent section, we direct our attention to individual differences in interpreting haptic messages. Given that participants employed varied standards or guidelines when designing haptic messages, they tended to adopt similar approaches when interpreting messages from others. We further investigated if there were discernible differences between interpretations of their own messages versus those from others. Figure 6 illustrates the accuracy rate for each individual across 3 distinct categories (full data set can be referenced in the Appendix D). On average, each individual interpreted an average of 14.88 calibrated haptic messages from others with an accuracy rate of valence $\mu = 0.25$, $\sigma = 0.10$ and arousal $\mu$ = 0.32, $\sigma$ = 0.14; 14.89 non-calibrated haptic messages from others with an accuracy rate of valence $\mu$ = 0.25, $\sigma$ = 0.09 and arousal $\mu$ = 0.32, $\sigma$ = 0.09; 3 haptic messages that they generated themselves with an accuracy rate of valence $\mu$ = 0.33, $\sigma$ = 0.26 and arousal $\mu$ = 0.33, $\sigma$ = 0.21. From this figure, it is evident that some participants vary notably from the average accuracy rates. When considering non-calibrated messages, both P4 and P8 have lower accuracy in valence, and P2 stands out with a reduced rate in arousal. Similarly, P3 and P18 exhibit challenges in interpreting the valence of calibrated messages originating from others, while P2, P17, and P18 show difficulty with arousal. Regarding non-calibrated messages from others, P4, P8, and P11 have lower accuracy rates for valence, and P17 has a markedly reduced accuracy rate for arousal in these messages. Given the considerable variations in accuracy rates of "messages from self", it remains inconclusive at this point whether participants exhibit an advantage in interpreting their own messages.

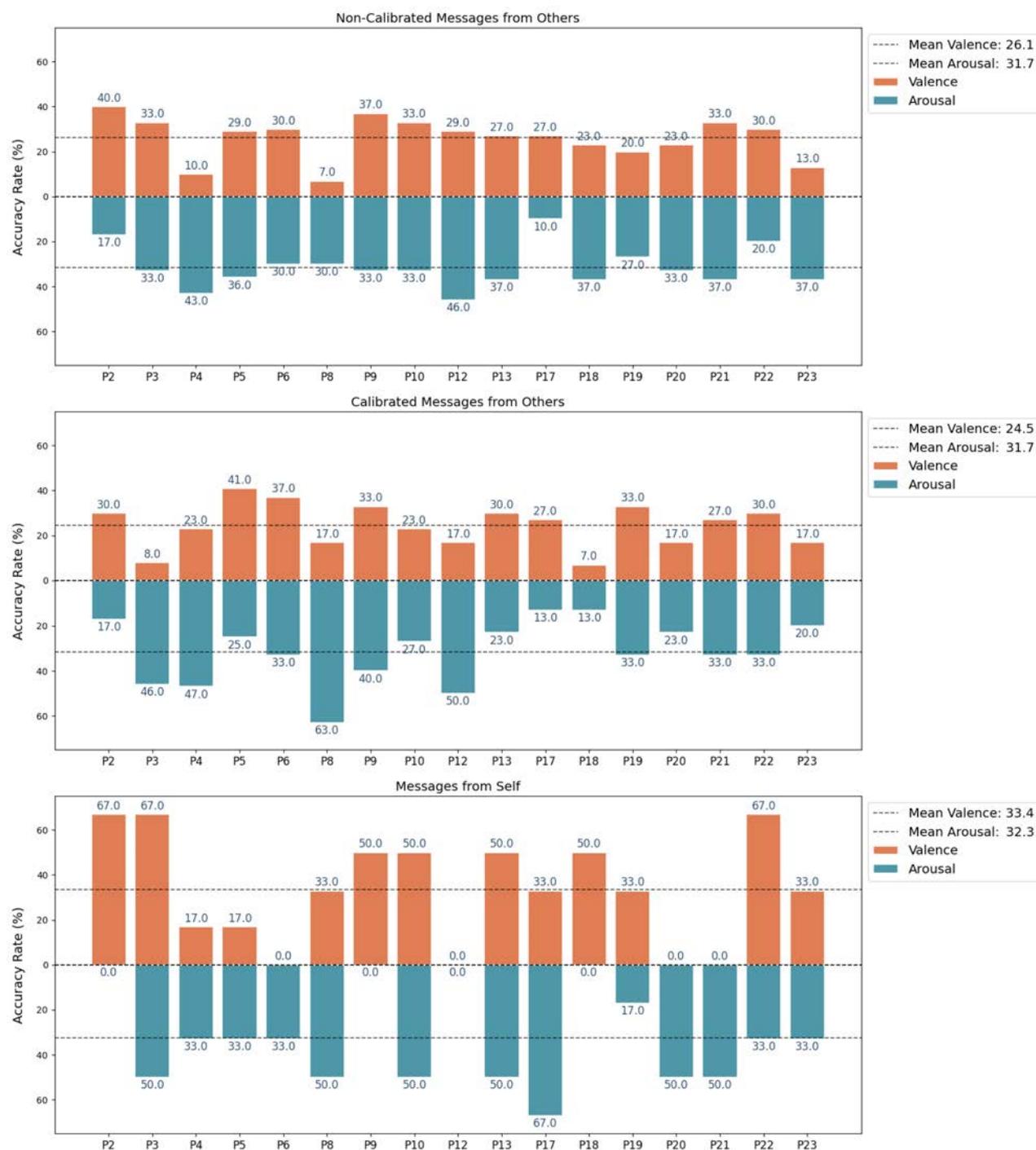

Figure 6: An overview of individual accuracy in interpreting calibrated haptic messages received from others, non-calibrated haptic messages received from others, and haptic messages originating from the participants themselves. The figure highlights performances that significantly deviate from the average, with some scoring notably above and others below it.

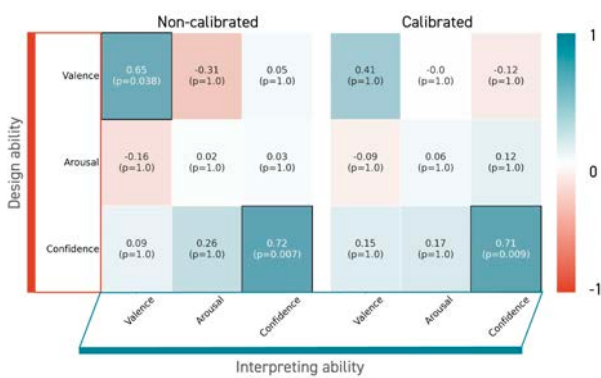

Figure 7: A correlation matrix between the proficiency in designing and interpreting calibrated (C) or non-calibrated (NC) haptic messages. "Design ability" refers to how well (i.e. accuracy rate of ) one's own haptic messages are correctly understood by others, while "interpretation ability" measures the accuracy rate of understanding haptic messages created by others. The primary focus is on the outcomes within the same category: for non-calibrated haptic messages, there is a positive correlation between the accuracy of valence interpretation and design proficiency ($\rho$ = 0.65, p < .05).

### 4.5 Haptic Empathy: Proficiency in Haptic Message Design

The interpretation of haptic messages is just a facet of the broader concept of haptic empathy. Another important component is the ability to design effective haptic messages. To evaluate participants' proficiency in this aspect, we measured the average accuracy rate of the haptic messages they crafted. Specifically, we computed the average percentage of messages designed by each participant that were interpreted correctly by others. This provides a more comprehensive understanding of participants' haptic empathy, encompassing both their interpretative and design capabilities. To investigate the possible relationship between these two abilities, we employed a Spearman's correlation and visualized the associations with a correlation matrix using a heatmap (see Fig.7). The Bonferroni adjustment ($\alpha$ = 0.05) was applied to control the family-wise error rate across 15 comparisons. Statistically significant results ($p_\text{adjusted}$ < 0.05) after correction are highlighted with a black stroke to emphasize the significant relationships. Given the categorizations of haptic messages into calibrated and non-calibrated, direct comparisons between these two groups are not meaningful. Our emphasis is on comparing outcomes within the same category. Additionally, since both "distance" and "accuracy" serve as measures of capabilities, albeit with inverse implications—where a reduced "distance" is preferable and a higher "accuracy" is indicative of better performance— we avoid directly contrasting these two metrics. Instead, we use "accuracy" for clarity and to facilitate intuitive visualization. For non-calibrated haptic messages, a positive correlation was found between the accuracy of valence interpretation and design proficiency ($\rho$ = 0.65, p < .05). Additionally, participants consistently demonstrated confidence across both design and interpretation phases, with strong positive correlations between the confidence levels of interpreting and designing non-calibrated messages ($\rho$ = 0.72, p < .001), as well as between the confidence levels for interpreting and designing calibrated messages ($\rho$ = 0.71, p < .001). This emphasizes the intrinsic link between design and interpretative abilities in the context of haptic messages.

To investigate the correlation between haptic empathy and other emotion-related abilities, we employed Spearman's correlation for these factors. Given the large number of comparisons (171), the Benjamini-Hochberg [6] procedure was applied to control the False Discovery Rate (FDR). Statistically significant results ($p_\text{adjusted}$ < 0.05) after correction are highlighted in Fig. 8 with a red stroke to emphasize the significant relationships. Due to the small sample size and numerous comparisons, we focus on the effect size rather than the p-value, as small sample sizes often lack the statistical power to detect significant relationships [61]. From the general ability aspect, there were a strong negative correlation between interpersonal EC and valence accuracy for interpreting calibrated messages ($\rho$ = -0.68, p < .05). Although the p-value for the correlation between Global EC and valence accuracy (p=0.134) was not statistically significant after the correction, a moderate negative correlation was noted ($\rho$ = -0.54). Additionally, negative correlations were noted between AIM and valence accuracy for designing ($\rho$ = -0.41, p = .297) and interpreting ($\rho$ = -0.49, p = .200) calibrated messages.

Delving deeper into the subscales, the "identifying" subscale at both the intrapersonal ($\rho$ = -0.48, p = .200) and interpersonal ($\rho$ = -0.46, p = .223) levels negatively correlated with the confidence level in interpreting calibrated messages. Similarly, the interpersonal "identifying" subscale negatively correlated with the confidence level in designing non-calibrated ($\rho$ = -0.45, p = .264) and calibrated messages ($\rho$ = -0.45, p = .232). The interpersonal "regulating" subscale negatively correlated with the valence accuracy of interpreting calibrated messages ($\rho$ = -0.48, p = .200).

### 4.6 Qualitative Analysis for Design Phase

After each study, we conducted semi-structured interviews with 23 participants (P2 to P24). Following the thematic analysis method by

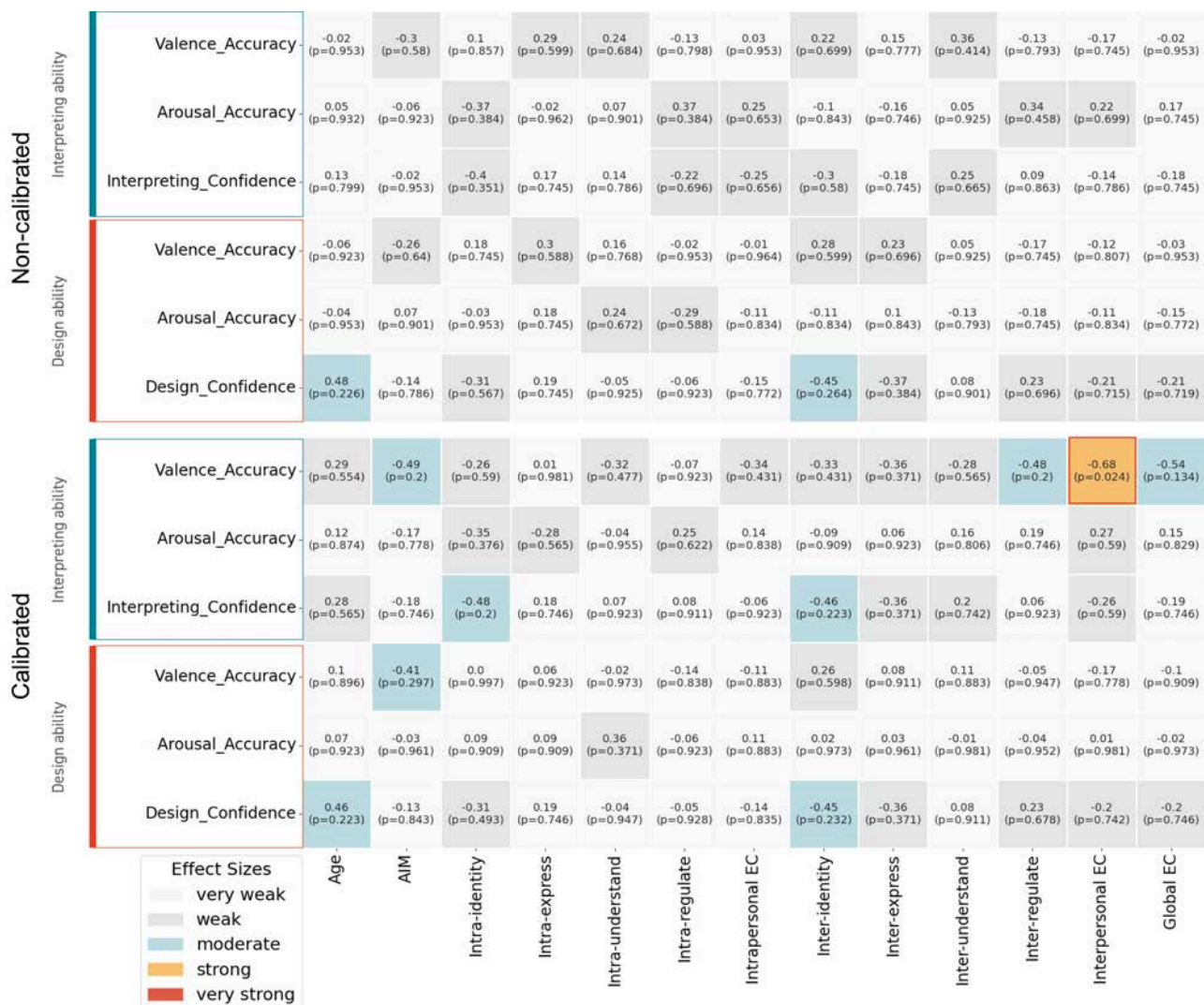

Figure 8: A correlation matrix illustrates the relationship between emotion-related abilities and haptic empathy. Cells are colored based on the effect size, with categories defined as follows: *very weak* ($0.0 \leq r \leq 0.19$, white), *weak* ($0.2 \leq r \leq 0.39$, light gray), *moderate* ($0.4 \leq r \leq 0.59$, blue), *strong* ($0.6 \leq r \leq 0.79$, yellow), and *very strong* ($0.8 \leq r \leq 1.0$, red). Significant results ($p < .05$) are highlighted with a red stroke, with one significant finding: interpersonal emotion comprehension (EC) is negatively correlated with valence accuracy from calibrated messages ($\rho = -0.68, p < .024$).

Maguire and Delahunt [53], we coded the transcripts into themes. The first author led the analysis, with themes reviewed and verified by the second author. Participants were asked: 1) how they designed haptic messages, 2) which tasks were easiest and most challenging, and why, and 3) any feedback on the experiment. From 159 data points and 41 codes, we refined 19 codes into four main themes. To facilitate discussion, we categorized film clips by participant descriptions: Disgust, Violent, Erotic, Surprise/Joy, Anger, Sadness, and Calm.

*4.6.1 Perceptive expression.* This theme outlines the three primary strategies participants employed to directly translate the visual cues from movie clips into haptic messages: drawing from motion, texture, and imagined sound. First of all, in total 6 people mentioned that they were trying to *"imitate the motion from the video"* (P7). To be more specific, P10 pointed out that he *"considered the amplitude of their movements and the rhythm from the movie clips"*. Similarly, P5 mentioned that he *"referenced the demo video and produced a haptic experience according to the intensity and frequency of the action"*. It is worth noting that P3 mentioned the distinction between left hand and right hand, *"because a lot of the film clips had to do with people interacting with different people or interacting with each other. Right side (hand), left side (hand), being the relationship or the conversation, or the interaction between two or more factors"* . We found in our observations that most participants did the same thing subconsciously, but only a few of them were aware of this tendency and explicitly pointed it out. Besides that, when designing haptic messages through this method, participants generally agreed that they were able to discern and design the difference in the movie clips containing *"significant actions"* (P5) such as erotic and violent movie clips. In contrast, intimacy-related clips or movie clips containing more nuanced emotions like sad or angry *"are more challenging, because it is hard to determine which specific action, given the many unique details, should be expressed through vibrations"* (P5).

From the visual information obtained from the video, participants not only utilized the motion but also made use of the tactile information. While designing haptic messages, participants (P9 and P20) mentioned that they *"based it on both rhythm and texture"*, with the calm film clip seeing the most of this approach. To be more specific, P20 described how he designed a film clip of calm as *"touched the clothing with one hand and the table with the other, representing the two distinct materials in the video. I used a softer object to represent the snow, and the other hand to convey the ice. Since its movement is relatively slow and lacks any distinct rhythm, I just kept moving continuously"*. P16 said she thought about the texture when designing haptic messages for erotic film clips, conveying the feel of skin by slowly touching her own arms. P22, who also used the texture of skin, tried to express disgust by clenching her palms and rubbing them slightly to recreate the feel of the gut that appeared in the film clip.

The last approach under this theme is referencing "sound" from the film clips. To minimize the interference of sound during the design phase, we did not use audio throughout, but participants could naturally add to this sensory information by imagination. P13, P15 and P23 mentioned that they can *"imagine some sounds based on their actions. It's like converting the visuals into the sounds I imagined, then transforming them into vibrations. More intense emotions are easier to express, such as slashing someone with a knife (violent). But when someone feels very happy or surprised, or when they feel very sad, it's hard to convey"*. This view is in agreement with participants who referenced actions, which was expected; after all, sound imagery is inherently movement-based. P21 also referenced sound, but he applied it to calm by *"simulating the cracking sound of the ice"*. This is similar to the participants who referenced the texture but focused on a different emphasis to deliver the features of the film clip.

*4.6.2 Empathetic expression.* This theme summarized the three methods participants used to represent their momentary emotional states through haptic messages: recreating emotional arousal, referencing physiological reactions, and perspective-taking (empathy). Regarding the first approach, representing emotional arousal, the outcome is similar to reproducing the actions in the film clips. Five participants believed that emotions with high arousal are relatively easier to represent, such as intense emotions like anger and erotic. A subtle distinction is that P19 and P24 pointed out that *"the erotic clips and the horror (violent) clips are different kinds of intensity and I tried to capture that"*. This implied that an additional dimension of emotion was recognized and considered during the design phase. While the above demonstrates a further enhancement in the participants' interpretation of clips, which involves emotional connotations, it still focuses on interpreting rather than based on their own feelings.

On the contrary, the subsequent design process incorporates more of the participants' own emotional expression. During the interviews, seven individuals mentioned their physical responses to the stimuli. Quote from P8, she had *"an immediate physical reaction after watching these videos... adapted to that without thinking much... and go with the flow of what she would behave. For example, for erotic ones, either move more excitedly or tap the bodies more; she would try to move away from my body for violent ones*. P4 also vividly described how he designed the message for violent, seeing it as an *"internally screaming"*, and then tried to *"express panic though that"*. Due to expressing physical reactions, most participants pointed out that *"other than calm and daily non-propositional ones contain more emotions, which it was easier to design"* (P16). P23 further mentioned that *"calm emotions are difficult to express because it's hard to get into this emotion immediately. Whereas if it was like a full-length movie, then I would be able to be much more immersed and feel the emotions"*. Not everyone agrees with this idea, P13 noted that *"calm relatively easy to convey by turning my emotions into gestures"*, which in turn can be transformed into vibrations.

Going a step further, six participants described that their design decisions were predicated on *"how it would feel if they were in that situation"* (P8, P12). This process aligns with the concept of empathy, which is defined as the attempt to comprehend the positive and negative experiences of another person without prejudice [96], by experiencing and understanding another person's affective or psychological state (i.e., put oneself in another person's shoes)[22, 100]. Given the variability in empathic abilities across individuals, participants reported a noticeable divergence in their perceived task difficulty. For example, P8 pointed out that *"the violent ones were more difficult because I don't really get involved in that much violence, which is hard for me to empathize. While on the contrary, calm or interpersonal (erotic) ones are easier for me to relate, which it's easier for me to make it to make messages, too"*. P12 on the other hand thought that *"calmness scene was the hardest, as it didn't feature any humans to be empathized"*. P2 shared idea similar to P12, mentioning the method to be only applicable in scenarios with *"more emotional and subtler hand or arm movements"*. He also emphasized that he would *"think what was the person breathing like in that situation"*. P17 also indicated that he *"can empathize with certain emotions, such as joy and surprise, generally positive emotions. In contrast, understanding feelings of sadness requires a more extended period of thinking to comprehend why one might feel that way. Similarly, anger requires deeper contemplation"*.

The level of difficulty, in addition to being related to the ability to empathize with specific emotions, also correlates with the perceived complexity of the emotions. One example is that the majority believed that erotic scenarios were easy to empathize with and, thus easy to express. P13 opined that such emotions were challenging and complex to express because *"it might involve excitement, happiness, intensity (excitement), and calmness, which is hard to reach a satisfactory state (for haptic message designing)"*. Both P13 and P17 highlighted that the film clip of *"sad contained mixed of emotions rather than singular feeling. However, due to the absence of contextual background, it becomes challenging to empathize"*. Even more, P21 felt *"surprise, fear, disgust, sadness, anger and all those things"* from the violent clip, which is an aspect we had not anticipated. Beyond the complex emotions elicited from stimulus itself, P6 indicated that while he *"felt shock from the intense stimulus(violent), the calm from the resting state hadn't entirely dissipated. This led to a weird mixed of emotional states"*. All the views presented above underscore the gap between a singular information transmission channel and the complexity of emotions.

*4.6.3 Metaphorical expression.* This theme delineates three design methods that involve implication or extrapolation, specifically: association, "language" construction, and cultural reference. Regarding association, participants listed four different examples during the interviews, calmness was associated with rainwater, or heartbeat; disgust was linked to high-frequency noise; and happiness was related to music. For a more in-depth explanation of the above examples, starting with "calm", three participants (P4, P6 and P11) each designed a message using their familiar sensations to convey the feeling of *"calm me down"*. For P4, the representative sensation was a heartbeat; for P6, it was the sound of raindrops; P11 did not specify a particular association. P17 felt that *"chaotic sensations can induce feelings of disgust, drawing parallels to experiences such as chalk screeching on a blackboard"*. P17 believed that *"humans (including himself) often derive joy from rhythmic elements, such as music. Hence, expressed feelings of happiness through music patterns"*. Those associations were not present in the content of the stimuli but derived from their everyday life experiences. In other words, the visual stimuli from the film clips triggered participants' memories. They then projected these experiences onto haptic messages, attempting to share a similar emotional experience with others.

"language" construction was proposed by four participants separately, and the interpretation of "language" varied for each individual, with each attributing a unique meaning to it. P18 thought the setup *"is more like a conversation of body language"*, which is partially aligned with P6's idea. P6 said that the setup reminded him of a *"'dialogue in silence', an event where verbal communication is forbidden, and participants must rely solely on other means for communication"*. He referred to two distinct experiences: "dialogue in the silence", where participants are guided by deaf individuals, and "dialogue in the dark", where they are guided by blind individuals. He found it intriguing how, in this context, the typical sense of speech was removed, leaving only vibrations for communication. Not only that, P6 remarked that the entire process *"is like trying to learn a new language and play with it"*. He emphasized the roles of the left and right hands during the process, exploring the difference between using one hand versus two and speculating about its potential impact on people's communication. Furthermore, he tried to *"use two hands, each performing different actions and questioned how such actions would be translated in this new 'language'"*. For P18, he likened the act of designing haptic messages using two fingers to drumming in music. He stated that *"all designs are based on musical expression"*. This line of thought indicates that participants were trying to endow the tactile experience with more diverse forms of expression, aiming for a more nuanced conveyance of information.

Cultural reference offers a unique perspective, and only one participant raised this point. P22 noted that erotic films remind her of *"the Chinese term 'pā pā pā'"*. This term, which has evolved and widely spread on the internet recent years, is a neologism used colloquially to describe intimate relations between men and women. Apart from this instance, however, the extension of cultural background was not frequently observed throughout the study. This might be attributed to our selection of stimuli, which lacked distinctive cultural variations. However, the participant's insights showcased the potential of vibrotactile stimuli to convey messages imbued with cultural meaning.

*4.6.4 Feedback.* This theme summarizes the participants' feedback and opinions regarding the display and the whole experience. Starting from the feedback about the display, participants initially evaluated the precision of the system. Both P2 and P8 opined that *"it is a high-fidelity display"*. However, P20 expressed a different opinion. She found that *"what she wanted to convey and what she actually could express (especially for sadness) were somewhat lacking. The device excelled in capturing the intensity of vibrations when tapping, even the difference between different surfaces (materials). However, when trying to express sadness by forcefully scraping the table with hands, the feedback felt no different than touching cloth"*. This is due to the limitations of tactile-mediated touch, which diminishes the conveyance of force-related information during the process. P5 also expressed a desire to perceive the direction of the force. In other words, she hoped to experience spatiotemporal vibrotactile patterns rather than a single vibration motor on the skin. Another often-mentioned suggestion was the desire for *"a wider range of materials (texture) for exploration and utilization"* (P17 and P20). During the study, it was observed that participants attempted to explore various materials around them. This interaction approach, which didn't restrict the use of both hands, encouraged individuals to explore a wider range of design possibilities instinctively. Building on this, P17 suggested that using all fingers of both hands, rather than just two fingers, would be more natural.

Regarding the difficulty of designing haptic messages, although it varied for different emotions, almost everyone believed that overall, this was a challenging task. However, there was an interesting phenomenon: as time goes on and with more practice, some people's adaptability and creativity for a new task gradually increase. For example, P6 felt *"lost at the beginning, but later on at the end, (he) felt freer and more confident"*. P3 also noticed that the methods they used gradually became more diverse, evolving from action-based to more emotion-incorporated. This implied that while there might be a dependence on a single strategy initially, more strategies and methods would emerge over time. While many participants adapted and grew more confident with time, others struggled due to a combination of physical exhaustion and the subtle differences involved in differentiating emotions. Participant *"got into the mentality of getting a bit tired, the fatigue gave less incentive to try to express (myself) in a more accurate way"* (P4). P11 gradually found that he *"started losing sight of the difference between anger and fear"*.

# 5 Discussion
## 5.1 Strategies for haptic communication

We identified three key strategies participants used to communicate emotions via a touch-based haptic interface: perceptive expression, empathetic expression, and metaphorical expression.

*5.1.1 Perceptive expression.* Participants relied heavily on translating visual cues from the film clips into tactile experiences. Even with only visual information provided, participants intuitively used elements such as movement and texture from the clips to express emotions through haptic means. This translation from visual to tactile cues highlights the role of cross-modal correspondences in sensory experiences, as discussed in previous research [51, 54]. While some visual cues were easily mapped to haptic outputs, others required more sensory imagination, indicating the subjective nature of this process. This subjectivity suggests the need for standardized guidelines to reduce variance and improve the effectiveness of haptic communication.

*5.1.2 Empathetic expression.* This strategy focused on how participants expressed their emotional states through haptic messages. Some participants mirrored the emotional arousal from the film clips, particularly when emotions were intense and easy to represent. Others based their designs on their physiological reactions to the stimuli. The variability in emotional responses, influenced by personal experiences, suggests that the same haptic message can evoke different interpretations. This diversity underscores the importance of personalizing tactile feedback to align with individual emotional thresholds and emphasizes the complexity of using haptic technology for emotional communication.

*5.1.3 Metaphorical expression.* In this strategy, participants used associations, "language" construction, and cultural references to convey emotions. As participants became more familiar with the design task, they adopted these more abstract methods to communicate emotions. The evolving nature of participants' strategies reflects their adaptability and creativity in using haptic interfaces, highlighting the potential to develop more nuanced forms of emotional communication through touch. Incorporating experiences from individuals with heightened tactile sensitivity, such as those with hearing impairments, could further inform the design of effective haptic systems.

Overall, during the initial design and encoding of haptic messages, standardized guidelines are crucial to establish a baseline of common knowledge and understanding. This ensures that the foundational elements of haptic communication, such as intensity, rhythm, or spatial distribution, remain consistent and interpretable across the majority of users and contexts. In contrast, customization or "calibration" is more relevant for the end-user experience. At this stage, the focus shifts to tailoring the tactile feedback to individual preferences and emotional thresholds. This allows for a more personalized and emotionally resonant interaction, ensuring that the standardized messages are adjusted to suit the unique needs and sensitivities of the user. The tactile feedback can be fine-tuned to align with the unique emotional sensitivities and preferences of the individual user, ensuring that the haptic messages resonate on a personal level. In summary, while standardization serves as the groundwork for broad usability, personalization enhances the nuanced effectiveness of the communication for the end-user. These guidelines address different stages of the process and are not in conflict but rather complementary in achieving effective haptic communication.

## 5.2 Emotion-related individual differences

The study examined the relationship between emotional traits and the ability to interpret haptic messages. Specifically, we focused on the subscales of identifying and regulating competence within emotional competencies and affect intensity.

*5.2.1 Identifying competence.* Emotional competencies are best understood as consistent traits rather than just knowledge or abilities [9]. Identifying competence refers to how well individuals recognize their own emotions through physical arousal, inner feelings, and thoughts. Over time, they also become better at discerning the emotions of others by observing external cues [42]. In this study, we investigated whether identifying competence (both intrapersonal and interpersonal) influences confidence in designing and interpreting haptic messages. Although our post-hoc analysis revealed no statistically significant correlation, initial descriptive trends suggested a potential negative relationship. Some participants (P6, P13, and P17), with high identifying competence, qualitatively reported experiencing multiple emotions simultaneously, which made it challenging to express these nuances through haptic messages. This heightened awareness of emotional complexity could lead to hesitation when trying to pin down a single emotional interpretation. Additionally, individuals with stronger emotional perception may also be more conscious of cultural, contextual, or personal differences in emotional expression, adding further complexity. Given the abstract nature of haptics as a medium for affective communication, this complexity may increase cognitive load, making it harder to decode unfamiliar stimuli.

*5.2.2 Regulating competence.* Emotional regulation, as defined by Mayer and Salovey [55], involves the ability to enhance, diminish, or modify one's own and others' emotions. Individuals skilled in regulation tend to better accept and manage both positive and negative emotions. Over time, they develop the capacity to either connect with or distance themselves from emotions, depending on their utility. While our post-hoc analysis did not reveal a statistically significant relationship, a non-significant trend suggested that individuals with higher interpersonal regulating competence showed a tendency toward lower accuracy in interpreting valence in calibrated haptic messages. One possible explanation is that individuals proficient in emotional regulation may have a tendency to overestimate emotional valence, leading to a wider gap between the intended and interpreted message. Their habitual regulation of emotional valence could interfere with accurately interpreting the subtleties of emotional stimuli conveyed through haptics.

*5.2.3 Affect Intensity.* Affect intensity refers to how strongly individuals typically experience their emotional responses [48]. According to arousal regulation theory, individuals seek an optimal level of arousal for task performance[16]. However, they differ in their baseline arousal and reactivity to stimuli. Larsen's research [48] indicates that individuals with high affect intensity tend to respond less strongly to sensory stimuli and often desire more intense arousal. In our study, this tendency may have contributed to lower accuracy in interpreting vibrotactile stimuli for individuals with high affect intensity, as these stimuli might not have provided the stronger arousal they sought.

Conversely, individuals with low affect intensity might perform better with vibrotactile stimuli because they require less stimulation to reach optimal arousal levels. This finding aligns with the idea that individuals with low affect intensity often feel that intense emotions disrupt their performance, while those with high affect intensity believe that emotional stimulation enhances their abilities. The relatively subtle nature of vibrotactile stimuli might have been less effective for individuals with high affect intensity, explaining the non-significant trends of negative relationship with accuracy in interpreting valence. However, as these trends were not statistically significant, they should be interpreted with caution. Future research could explore whether individual differences in affect intensity systematically influence haptic emotion recognition. Examining these relationships with a larger sample or more diverse haptic stimuli may provide further insights into the role of affect intensity in vibrotactile perception.

### 5.3 Design guidelines for affective haptic interface

In our study, we collected 187 haptic messages from 24 participants, each created based on one of four types of film clips (HVHA, HVLA, LVHA, and LVLA). From these, 72 haptic messages were selected and interpreted by 19 participants, producing 593 samples. The overall interpretation accuracy ranged from 24-38%, with LVHA accuracy falling below the chance level (0.21). This aligns with the findings of Rantala et al. [68], which showed that unpleasant, aroused emotions are less effectively conveyed through finger touch compared to squeezing. Their research suggests that squeezing is more effective for unpleasant emotions, while finger touch is better for pleasant, relaxed emotions. This highlights the importance of designing haptic systems with specific gestures tailored to distinct emotions.

Our study's lower interpretation accuracy compared to other research (46.9%–71.4%) [21] can be attributed to differences in methodology. We used 9-point SAM scales, which added nuance and complexity to the interpretation process, rather than simple emotional labels. Furthermore, our haptic messages were vibration-based, yet participants demonstrated remarkable creativity by incorporating elements beyond intensity and rhythm, such as texture and cultural context. This feedback underscores the need for a broader range of tactile feedback, suggesting the inclusion of diverse materials and force feedback to enhance emotional expression. Multiple actuation technologies should be integrated into future haptic systems to offer richer, more varied feedback.

We also implemented a calibration system to account for individual skin sensitivities. Each participant interpreted haptic messages twice: once in a calibrated form and once non-calibrated. However, chi-square analysis indicated no significant differences in accuracy between the two formats, suggesting that calibration did not substantially impact overall interpretation accuracy. Nonetheless, individual analyses indicated that calibration could influence specific subsets of participants.

Based on these findings, we propose five key guidelines for developing haptic interfaces that effectively convey empathy:

**Incorporate Crossmodal Correspondences for Enhanced Emotional Communication:** Participants demonstrated a natural ability to translate visual cues into tactile sensations, particularly through movement and texture. This crossmodal correspondence, where sensory experiences from different modalities are intuitively linked, is critical for designing haptic messages that effectively communicate emotions. When designing haptic interfaces for empathy, it is essential to map emotional cues from other sensory modalities (e.g., visual or auditory) to corresponding haptic stimuli, such as vibration intensity, rhythm, or texture, to facilitate clearer emotional communication [51, 54].

**Design for Personalization and Flexibility in Emotional Thresholds:** The study revealed significant variability in how participants designed and interpreted haptic messages, influenced by their individual emotional thresholds and personal experiences. Some participants struggled to express or interpret certain emotions, particularly when their personal emotional threshold was higher or lower than the intensity of the stimulus. Therefore, haptic interfaces must be adaptable, allowing for personalization based on the user's emotional thresholds to ensure accurate emotional communication across a diverse set of users [42]. A system that can calibrate to individual preferences and emotional intensities will better support the nuanced nature of emotional expression.

**Leverage Physiological Reactions to Enhance Authenticity of Emotional Expression:** Participants who designed haptic messages based on their own physiological reactions, such as heart rate or muscle tension, found this method intuitive and effective for conveying emotions. This suggests that integrating physiological feedback into haptic design, such as heart rate or skin conductance, could enhance the authenticity of emotional messages. Future haptic systems should explore real-time physiological data to adjust haptic feedback dynamically, providing more personalized and emotionally resonant experiences for users [55].

**Utilize Metaphors and Cultural References to Enrich Emotional Communication:** As participants grew more familiar with the design task, they began to use metaphorical expressions and cultural references to enhance the emotional depth of their haptic messages. This shows the importance of allowing users to incorporate abstract or metaphorical concepts into haptic feedback. Haptic interfaces should provide a flexible design space where users can express complex emotions through metaphors, such as associating happiness with rhythmic patterns or sadness with slow, undulating vibrations. Including culturally relevant references in the design process can further enhance the effectiveness of haptic communication across diverse user groups [9].

**Standardize Iterative Haptic Encoding Strategies:** Participants in the study expressed growing confidence and creativity in designing haptic messages as they engaged more with the task. This suggests that providing users with opportunities for iterative design and feedback can significantly improve their ability to communicate emotions through haptics. Haptic systems should incorporate mechanisms that allow users to test, refine, and receive feedback on their haptic messages. Such iterative processes will help users hone their haptic empathy and develop a more intuitive understanding of how to convey emotions through touch [40, 68].

## 6 Limitations and Future Work

Our haptic display was portable, user-friendly, and granted participants considerable creative freedom in designing haptic messages. In terms of fidelity, it is considered as a high-fidelity device. However, based on user feedback, there are areas for improvement. For haptic recordings, improving user interactions, such as allowing the use of all five fingers instead of just two, would mimic a more natural experience. Although the current design approach for haptic messages offers considerable freedom, it can be challenging to control. Inviting participants to join the haptic message editing process might strike a better balance between freedom and control.

Aside from the limitations of the haptic display, we emphasize that this research exclusively utilized vibrotactile actuation, which is only a fraction of what our somatosensory system can perceive. Adding to this, the stimuli presented to users in the study focused solely on video; stimuli that directly engage the other senses in the real world could alter user behavior and provide different insights and results. Admittedly, for this iteration, while some valuable insights were gained on how users design and interpret haptic messages, the small sample size and the lower degree of control over participants' emotional regulation raised some caveats regarding the applicability of the study. Future iterations will seek to address these issues, as well as incorporate additional haptic modalities for the hardware. We will explore the outcomes of incorporating additional haptic modalities in our future works.

Investigating individual differences in the design and interpretation of affective haptic messages presented great challenges. As mentioned in the discussion, users also incorporated aspects of cultural context into their design, which introduced further complications. This was because our participants came from multiple nationalities and cultural backgrounds, and such culture-specific elements would have been "lost in translation," contributing to a lower accuracy rate during the interpretation phase. Additionally, the relatively small sample size limited the statistical power of our findings, making it difficult to draw definitive conclusions about the relationships between emotional traits and haptic message comprehension. A number of variables were considered during the analysis, yet conceptualizing the ability to convey and comprehend emotional meanings through vibrotactile stimulation as "Haptic Empathy" raises further questions about the nature of this phenomenon. These factors not only affect the design of practical applications but also present avenues for future research.

## 7 Conclusion

This study yielded a total of 187 haptic messages from 24 participants, with a grand total of 593 samples including the interpretation data. While our analysis revealed that individuals are capable of effectively communicating certain emotional information through haptic messages when in distinct emotional states, the standards for this communication remain highly individualistic. Additionally, correlations between proficiency in designing and interpreting haptic messages were explored; however, no conclusive evidence was found linking these abilities to specific emotional traits. Although our findings do not establish a definitive relationship between interpretation/design ability and emotional traits, they raise important questions about the role of emotional ability in affective haptic design. This user-centered study seeks to incorporate the concept of haptic empathy into the broader framework of emotional intelligence, emphasizing the significance of individual differences in mediated social touch.

Furthermore, by delving into user understanding, we are creating a framework and outlining potential future research paths that focus on individual variations in the mediated social touch. While we have not conducted evaluations for the suggestions presented in this paper, we hope this overview will offer useful insights for enhancing the naturalness of affective haptic interactions.

## Acknowledgments

This work was supported by the JST Moonshot R&D Program "Cybernetic being" Project (Grant number JPMJMS2013), JST SPRING (Grant number JPMJSP2123), the University of Auckland Faculty of Science Research Development Fund (Grant number 3731533), the Japan Society for the Promotion of Science, the Inamori Research Institute for Science and NTT DOCOMO, INC.

# A Appendix: Snapshots of the clips included in the haptic content video.

Figure 9 shows snapshots of the different video clips that were included in the haptic content video. The 3-minute-long video was shown to participants to familiarize them with the haptic presentation device and how it worked by showcasing several video clips with pre-designed haptic feedback that would simulate the real-life haptic feedback one would get in the situations portrayed in the clips. Participants were asked to hold each hemisphere of the device separately and imitate the movements in the video for further immersion.

# B Appendix: Self-assessment form for the design phase and interpretation phase

Figure 10 presents the questionnaires used in both the design and interpretation phases to assess valence and arousal levels.

# C Appendix: the relationship between video type and SAM accuracy (valence accuracy and arousal accuracy)

To investigate the relationship between video type and SAM accuracy (valence accuracy and arousal accuracy), we conducted a chi-square test of independence. There was a weak association between film clip types and valence accuracy, $X^2(6, N = 650) =$ 38.606, $p < .001$. The difference between "incorrect" and "partially correct" was statistically significant for the four film clip types, $X^2(3, N = 564) = 29.239, p < .001$. The difference between "incorrect" and "correct" was statistically significant for the four video types, $X^2(3, N = 479) = 12.931, p = .005$. The frequency distribution of valence accuracy in Table 5 shows how valence accuracy is distributed between "incorrect", "partially correct" and "correct".

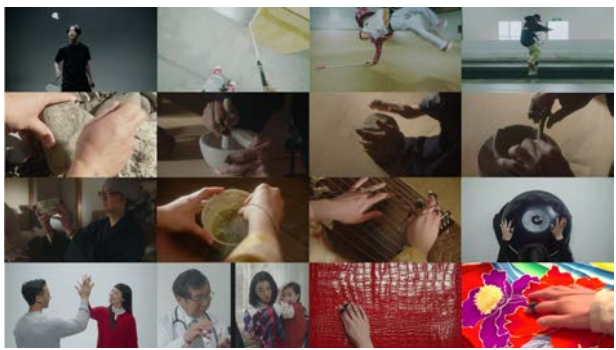

Figure 9: Snapshots of the clips included in the haptic content video. Each of the video clips contained contains demos of different haptic sensations in various contexts. These clips, arranged from the upper left to the lower right, showcase a range of activities: playing badminton, a visually impaired man walking, a visually impaired man dancing, a visually impaired man skateboarding, knocking stones, grinding clay (pottery), wedging (pottery), shaping (ceramics), shaping clay, tea making, Japanese koto, handpan, high five, telemedicine, texture of leather and texture of fabric.

There was no significant association between film clip types and arousal accuracy, $X^2(6, N = 650) = 5.885, p = 0.436$. Similarly, no significant association was found between "incorrect" and "partially correct" responses across the four film clip types, $X^2(3, N = 532) = 5.165, p = 0.160$, nor between "incorrect" and "correct" responses, $X^2(3, N = 468) = 2.105, p = 0.551$.

## D Appendix: Individual Accuracy Rate

Tables 6 and 7 present a detailed breakdown of each individual's accuracy rate when interpreting different types of haptic messages. The results include accuracy rates for all calibrated messages, calibrated messages from others, all non-calibrated messages, non-calibrated messages from others, and messages from the individuals themselves. Blank cells indicate that the individual did not interpret that specific type of message.

## E Appendix: Chance Levels

The following outlines the detailed mathematical framework for calculating the chance level in this study.

### E.1 Correct (Valence Distance = 0)

For a valence distance of 0, the interpreter's score must exactly match the designer's score. Since there are 9 possible scores, the probability of this occurring is: $P(\text{Correct}) = \frac{1}{9} \approx 0.111$ (11.1%)

Figure 10: Self-assessment form for the design phase and interpretation phase. Self-assessment manikins (SAM) for valence and arousal were employed for both phases. It also includes preferences, familiarity, and confidence in the design phase; tactile description and confidence in the interpretation phase.

### E.2 Partially Correct (Valence Distance = 1)

For a valence distance of 1, the interpreter's score must differ from the designer's score by exactly one unit. The number of possible neighbors depends on the designer's score:

- Scores 2 through 8 have 2 valid neighbors (e.g., score 2 has neighbors 1 and 3).
- Scores 1 and 9 (endpoints) have only 1 valid neighbor each (e.g., score 1 has neighbor 2).

The probabilities are calculated as follows:

Table 5: Crosstabulation of film clip types and valence accuracy. The table displays the count of "incorrect", "partially correct", and "correct" valence interpretations for each video type (LVHA, HVHA, HVLA, LVLA), with their respective expected counts and percentages within each video type category. The chance levels for being "incorrect", "partially correct", and "correct" are 69.1%, 19.8%, and 11.1%, respectively. The LVLA video type has the highest interpretation accuracy at 54.4%, with all results exceeding the expected chance level.

|  |  |  | Valence accuracy | | | |
|---|---|---|---|---|---|---|
|  |  |  | Incorrect | Partially correct | Correct | Total |
| Video Type | LVHA | Count | 124 | 28 | 13 | 165 |
|  |  | % within Video Type | 75.2% | 17.0% | 7.9% | 100% |
|  |  | % within Valence Accuracy | 31.6% | 16.4% | 15.1% | 25.4% |
|  |  | % of total | 19.1% | 4.3% | 2.0% | 25.4% |
|  | HVHA | Count | 102 | 42 | 33 | 177 |
|  |  | % within Video Type | 57.6% | 23.7% | 18.6% | 100% |
|  |  | % within Valence Accuracy | 6.0% | 24.6% | 38.4% | 27.2% |
|  |  | % of total | 15.7% | 6.5% | 5.1% | 27.2% |
|  | HVLA | Count | 104 | 42 | 20 | 166 |
|  |  | % within Video Type | 62.7% | 25.3% | 12.0% | 100% |
|  |  | % within Valence Accuracy | 26.5% | 24.6% | 23.3% | 25.5% |
|  |  | % of total | 16.0% | 6.5% | 3.1% | 25.5% |
|  | LVLA | Count | 63 | 59 | 20 | 142 |
|  |  | % within Video Type | 44.4% | 41.5% | 14.1% | 100% |
|  |  | % within Valence Accuracy | 16.0% | 34.5% | 23.3% | 21.8% |
|  |  | % of total | 9.7% | 9.1% | 3.1% | 21.8% |
| Total |  | Count | 393 | 171 | 86 | 650 |
|  |  | % within Video Type | 60.5% | 26.3% | 13.2% | 100% |
|  |  | % within Valence Accuracy | 100% | 100% | 100% | 100% |
|  |  | % of total | 60.5% | 26.3% | 13.2% | 100% |

**Scores 1 and 9:** Each score has only 1 neighbor, and the probability of selecting it randomly is: $P(\text{Partially Correct} \mid \text{Score 1 or 9}) = \frac{1}{9}$. Since there are 2 endpoint scores (1 and 9), their combined contribution is $P(\text{Endpoints}) = 2 \cdot \frac{1}{9} = \frac{2}{9}$.

**Scores 2 through 8:** Each score has 2 neighbors, and the probability of selecting one of them is: $P(\text{Partially Correct} \mid \text{Score 2 to 8}) = \frac{2}{9}$. Since there are 7 middle scores, their combined contribution is: $P(\text{Middle Scores}) = 7 \cdot \frac{2}{9} = \frac{14}{9}$.

**Total Probability:** Summing the contributions of the endpoints and middle scores: $P(\text{Partially Correct}) = \frac{2}{9} + \frac{14}{9} = \frac{16}{9} \approx 0.198\ (19.8\%)$.

### E.3 Incorrect (Valence Distance > 1)

For a valence distance greater than 1, the interpreter's score must be two or more units away from the designer's score. The total probability for this case is the complement of the probabilities for "correct" and "partially correct": $P(\text{Incorrect}) = 1 - P(\text{Correct}) - P(\text{Partially Correct})$.

Substituting the known probabilities: $P(\text{Incorrect}) = 1 - 0.111 - 0.198 = 0.691\ (69.1\%)$.

### E.4 Summary of Chance Levels

The chance levels for valence accuracy based on random agreement are as follows:

- Correct (Valence Distance = 0): 11.1%.
- Partially Correct (Valence Distance = 1): 19.8%.
- Incorrect (Valence Distance > 1): 69.1%.

Table 6: Individual accuracy rate of valence. The table presents the mean and standard deviation (±SD) of valence accuracy and distance measures for each participant under Calibration and Non-Calibration conditions. Results are further categorized based on the type of haptic message received: All 4 types (aggregated performance across all message types), Other's haptic message (responses to haptic messages from others), and Own haptic message (only in the Non-Calibration condition, referring to self-generated haptic messages).

| | Valence Mean(±SD) | | | | | | | | | |
|---|---|---|---|---|---|---|---|---|---|---|
| | Calibration | | | | Non-calibration | | | | | |
| | All 4 type | | Other's haptic message | | All 4 type | | Other's haptic message | | Own haptic message | |
| | Distance | Accuracy | Distance | Accuracy | Distance | Accuracy | Distance | Accuracy | Distance | Accuracy |
| P2 | 1.56±1.15 | 0.36±0.38 | 1.73±1.16 | 0.3±0.37 | 1.33±1.24 | 0.44±0.45 | 1.47±1.3 | 0.4±0.47 | 0.67±0.58 | 0.67±0.29 |
| P3 | 2.07±1.22 | 0.2±0.37 | 2.42±1.08 | 0.08±0.29 | 2±1.85 | 0.39±0.4 | 2.27±1.91 | 0.33±0.41 | 0.67±0.58 | 0.67±0.29 |
| P4 | 2.72±1.93 | 0.22±0.35 | 2.87±2.07 | 0.23±0.37 | 3.17±1.65 | 0.11±0.27 | 3.4±1.68 | 0.1±0.28 | 2±1 | 0.17±0.29 |
| P5 | 1.63±1.42 | 0.37±0.44 | 1.62±1.54 | 0.41±0.46 | 1.82±1.38 | 0.26±0.36 | 1.86±1.51 | 0.29±0.38 | 1.67±0.58 | 0.17±0.29 |
| P6 | 1.83±1.34 | 0.31±0.39 | 1.73±1.44 | 0.37±0.4 | 2.67±2 | 0.25±0.39 | 2.73±2.19 | 0.3±0.41 | 2.33±0.58 | 0±0 |
| P8 | 2.56±1.69 | 0.19±0.35 | 2.73±1.75 | 0.17±0.36 | 2.89±1.68 | 0.11±0.21 | 3.13±1.68 | 0.07±0.18 | 1.67±1.15 | 0.33±0.29 |
| P9 | 2.17±1.95 | 0.36±0.33 | 2.13±1.77 | 0.33±0.31 | 1.72±1.6 | 0.39±0.4 | 1.6±1.24 | 0.37±0.4 | 2.33±3.21 | 0.5±0.5 |
| P10 | 2.61±2.15 | 0.28±0.35 | 2.8±2.18 | 0.23±0.32 | 2.06±1.92 | 0.36±0.38 | 2.13±1.96 | 0.33±0.36 | 1.67±2.08 | 0.5±0.5 |
| P11 | 2±3.46 | 0.67±0.58 | | | 2.89±2.08 | 0.22±0.35 | 3.07±1.83 | 0.13±0.23 | 2±3.46 | 0.67±0.58 |
| P12 | 2.5±1.47 | 0.14±0.29 | 2.27±1.39 | 0.17±0.31 | 2.47±1.74 | 0.24±0.31 | 2.21±1.72 | 0.29±0.32 | 3.67±1.53 | 0±0 |
| P13 | 1.94±1.55 | 0.33±0.38 | 2.13±1.6 | 0.3±0.37 | 2.11±1.78 | 0.31±0.35 | 2.33±1.84 | 0.27±0.32 | 1±1 | 0.5±0.5 |
| P15 | 2.67±0.58 | 0±0 | | | 2.22±1.17 | 0.17±0.3 | 2.13±1.25 | 0.2±0.32 | 2.67±0.58 | 0±0 |
| P17 | 2.06±1.7 | 0.28±0.35 | 2±1.69 | 0.27±0.32 | 1.72±1.13 | 0.28±0.35 | 1.6±0.91 | 0.27±0.32 | 2.33±2.08 | 0.33±0.58 |
| P18 | 2.44±1.38 | 0.14±0.33 | 2.47±0.92 | 0.07±0.26 | 2.28±1.74 | 0.28±0.35 | 2.27±1.49 | 0.23±0.32 | 2.33±3.21 | 0.5±0.5 |
| P19 | 1.94±1.66 | 0.33±0.38 | 2±1.73 | 0.33±0.36 | 2.22±1.48 | 0.22±0.31 | 2.33±1.5 | 0.2±0.25 | 1.67±1.53 | 0.33±0.58 |
| P20 | 2.33±1.28 | 0.14±0.29 | 2.33±1.4 | 0.17±0.31 | 2.33±1.41 | 0.19±0.35 | 2.33±1.54 | 0.23±0.37 | 2.33±0.58 | 0±0 |
| P21 | 2.67±1.91 | 0.22±0.35 | 2.6±2.06 | 0.27±0.37 | 2.17±1.5 | 0.28±0.35 | 2±1.56 | 0.33±0.36 | 3±1 | 0±0 |
| P22 | 1.72±1.36 | 0.36±0.38 | 1.93±1.39 | 0.3±0.37 | 1.56±1.29 | 0.36±0.41 | 1.73±1.33 | 0.3±0.41 | 0.67±0.58 | 0.67±0.29 |
| P23 | 2.5±1.42 | 0.19±0.35 | 2.47±1.3 | 0.17±0.31 | 2.61±1.38 | 0.17±0.34 | 2.6±1.24 | 0.13±0.3 | 2.67±2.31 | 0.33±0.58 |
| Average | 2.19±1.66 | 0.27±0.36 | 2.24±1.66 | 0.26±0.36 | 2.29±1.69 | 0.26±0.36 | 2.34±1.7 | 0.25±0.35 | 1.96±1.65 | 0.33±0.4 |

Table 7: Individual accuracy rate of arousal. The table presents the mean and standard deviation (±SD) of arousal accuracy and distance measures for each participant under Calibration and Non-Calibration conditions. The results are further categorized based on the type of haptic message received: All 4 types (aggregated performance across all message types), Other's haptic message (responses to haptic messages from others), and Own haptic message (only in the Non-Calibration condition, referring to self-generated haptic messages).

| | Arousal Mean(±SD) | | | | | | | | | |
|---|---|---|---|---|---|---|---|---|---|---|
| | Calibration | | | | Non-calibration | | | | | |
| | All 4 type | | Other's haptic message | | All 4 type | | Other's haptic message | | Own haptic message | |
| | Distance | Accuracy | Distance | Accuracy | Distance | Accuracy | Distance | Accuracy | Distance | Accuracy |
| P2 | 2.56±1.34 | 0.14±0.29 | 2.6±1.45 | 0.17±0.31 | 2.39±1.14 | 0.14±0.23 | 2.4±1.24 | 0.17±0.24 | 2.33±0.58 | 0±0 |
| P3 | 1.4±1.3 | 0.47±0.35 | 1.5±1.45 | 0.46±0.4 | 1.94±1.66 | 0.36±0.41 | 2.13±1.77 | 0.33±0.45 | 1±0 | 0.5±0 |
| P4 | 1.67±1.68 | 0.44±0.45 | 1.47±1.51 | 0.47±0.44 | 1.89±2 | 0.42±0.43 | 1.73±1.94 | 0.43±0.42 | 2.67±2.52 | 0.33±0.58 |
| P5 | 2.42±1.89 | 0.26±0.39 | 2.56±1.97 | 0.25±0.37 | 1.82±1.51 | 0.35±0.42 | 1.86±1.56 | 0.36±0.41 | 1.67±1.53 | 0.33±0.58 |
| P6 | 1.89±1.68 | 0.33±0.38 | 2±1.81 | 0.33±0.41 | 1.94±1.51 | 0.31±0.39 | 2.07±1.62 | 0.3±0.41 | 1.33±0.58 | 0.33±0.29 |
| P8 | 1.28±1.74 | 0.61±0.44 | 1.13±1.6 | 0.63±0.44 | 1.67±1.28 | 0.33±0.34 | 1.6±0.99 | 0.3±0.32 | 2±2.65 | 0.5±0.5 |
| P9 | 1.83±1.5 | 0.33±0.38 | 1.33±0.98 | 0.4±0.39 | 2.11±1.6 | 0.28±0.39 | 1.67±1.29 | 0.33±0.41 | 4.33±1.15 | 0±0 |
| P10 | 2.5±2.26 | 0.31±0.39 | 2.8±2.37 | 0.27±0.42 | 1.94±1.89 | 0.36±0.38 | 2.13±2.03 | 0.33±0.41 | 1±0 | 0.5±0 |
| P11 | 1.33±1.53 | 0.5±0.5 | | | 1.83±1.72 | 0.39±0.44 | 1.93±1.79 | 0.37±0.44 | 1.33±1.53 | 0.5±0.5 |
| P12 | 1.61±1.5 | 0.42±0.43 | 1.4±1.5 | 0.5±0.42 | 1.71±1.61 | 0.38±0.42 | 1.5±1.65 | 0.46±0.41 | 2.67±1.15 | 0±0 |
| P13 | 1.83±1.2 | 0.28±0.39 | 1.87±1.06 | 0.23±0.37 | 1.67±1.57 | 0.39±0.47 | 1.67±1.54 | 0.37±0.48 | 1.67±2.08 | 0.5±0.5 |
| P15 | 2.67±2.89 | 0.33±0.29 | | | 2.17±2.04 | 0.36±0.38 | 2.07±1.94 | 0.37±0.4 | 2.67±2.89 | 0.33±0.29 |
| P17 | 2.61±1.88 | 0.22±0.35 | 3±1.81 | 0.13±0.3 | 2.67±1.88 | 0.19±0.35 | 3.07±1.79 | 0.1±0.28 | 0.67±0.58 | 0.67±0.29 |
| P18 | 2.39±1.2 | 0.11±0.32 | 2.27±1.22 | 0.13±0.35 | 1.94±1.51 | 0.31±0.42 | 1.73±1.53 | 0.37±0.44 | 3±1 | 0±0 |
| P19 | 2.22±1.7 | 0.31±0.39 | 2.13±1.77 | 0.33±0.41 | 2.28±1.67 | 0.25±0.35 | 2.2±1.74 | 0.27±0.37 | 2.67±1.53 | 0.17±0.29 |
| P20 | 1.94±1.51 | 0.28±0.35 | 2.13±1.6 | 0.23±0.37 | 1.61±1.29 | 0.36±0.38 | 1.73±1.39 | 0.33±0.41 | 1±0 | 0.5±0 |
| P21 | 1.44±1.04 | 0.36±0.41 | 1.53±1.06 | 0.33±0.41 | 1.61±1.5 | 0.39±0.37 | 1.73±1.58 | 0.37±0.35 | 1±1 | 0.5±0.5 |
| P22 | 2.22±1.86 | 0.33±0.42 | 2.33±1.95 | 0.33±0.41 | 2.5±1.62 | 0.22±0.35 | 2.67±1.63 | 0.2±0.32 | 1.67±1.53 | 0.33±0.58 |
| P23 | 2.28±1.49 | 0.22±0.31 | 2.4±1.55 | 0.2±0.32 | 1.89±1.6 | 0.36±0.33 | 1.93±1.71 | 0.37±0.35 | 1.67±1.15 | 0.33±0.29 |
| Average | 2.29±1.69 | 0.26±0.36 | 2.05±1.58 | 0.29±0.38 | 1.99±1.61 | 0.32±0.38 | 2.01±1.63 | 0.31±0.38 | 1.91±1.53 | 0.33±0.36 |